\documentclass[preprint,3p]{elsarticle}
\usepackage{amsmath}
\usepackage{amssymb}
\usepackage{enumerate}

\newtheorem{Theorem}{Theorem}
\newtheorem{Lemma}{Lemma}

\begin{document}
\title{Projective Structure and Holonomy in 4-dimensional Lorentz Manifolds}
\author[gsh]{Graham S.˜Hall}\ead{g.hall@abdn.ac.uk}
\author[dpl]{David P.˜Lonie\corref{cor1}}\ead{DLonie@aol.com}\cortext[cor1]{Corresponding author}
\address[gsh]{Institute of Mathematics, University of Aberdeen, Aberdeen, AB24 3UE, U.K.}
\address[dpl]{108e Anderson Drive, Aberdeen AB15 6BW, Scotland, UK}

\begin{abstract}
This paper studies the situation when two $4$-dimensional Lorentz manifolds (that is, space-times) admit the same
(unparametrised) geodesics, that is, when they are projectively
related. A review of some known results is given and then the
problem is considered further by treating each holonomy type in turn for the
space-time connection. It transpires that all holonomy
possibilities can be dealt with completely except the most general one and that
the consequences of two space-times being projectively related leads,
in many cases, to their associated Levi-Civita connections being
identical.
\end{abstract}


\maketitle
\section{Introduction and Notation}

There has been some recent interest in the study of projective
relatedness of (metric) connections and, in particular, within
Einstein's theory \cite{1,2,3,4,5,6,7,8,9}. Thus, roughly speaking, one
assumes that one has two Lorentz metrics on a given space-time $M$
whose Levi-Civita connections give rise to the same set of geodesic
paths (\emph{unparametrised} geodesics) on the space-time manifold
and then tries to find the relationship between these metrics and
connections. Such a problem has an immediate geometrical appeal but
is also of obvious interest within Einstein's general relativity
because of the Newton-Einstein principle of equivalence. This paper
continues the general study of finding those metrics $g$ whose
Levi-Civita connection $\nabla$ can be recovered uniquely from its
geodesics, that is, $\nabla$ is such that if any other Levi-Civita
connection $\nabla'$ of a Lorentz metric $g'$ on $M$ has the same
geodesic paths as those of $\nabla$ then $\nabla=\nabla'$. If this
is the case, the relationship between $g$ and $g'$ can be found from
holonomy theory. If $\nabla=\nabla'$ is not the conclusion
one can still compute $g'$ in terms of $g$ in all cases except the most
general holonomy case and even for this situation the solution for $g'$ can be achieved under certain conditions. A brief summary of some results in this direction
is given in sections 4 and 5 following preliminary work in the next
two sections. Sections 5 and 6 of this paper will concentrate on
a study which reveals a close relationship between the holonomy type
of a space-time and the varying degrees of ability of its associated
Levi-Civita connection to be recoverable from its geodesic paths.

Throughout this paper, $M$ will denote a 4-dimensional, Hausdorff,
connected, smooth manifold which admits a smooth metric $g$ of
Lorentz signature $(-,+,+,+)$. The pair $(M,g)$ is called a {\em
space-time}.  All structures naturally occurring on $M$ will be
assumed smooth. The unique, symmetric Levi-Civita connection arising
on $M$ through $g$ is denoted by $\nabla$ and, in a coordinate
domain of $M$, its Christoffel symbols are written $\Gamma^a_{bc}$.
The type $(1,3)$ curvature tensor associated with $\nabla$ is
denoted by $Riem$ and its (coordinate) components are written
$R^a_{\ bcd}$. The Ricci tensor, $Ricc$, derived from $Riem$, has
components $R_{ab}\equiv R^c_{\ acb}$ and $R\equiv R_{ab}g^{ab}$ is the Ricci
scalar.

For $m\in M$, $T_mM$ denotes the {\em tangent space to $M$ at $m$}
and this will, for convenience, be identified with the cotangent
space to $M$ at $m$, through the isomorphism arising from the metric $g(m)$ by index raising and
lowering. If no ambiguity could arise, the same symbol will be used
for both a vector (field) and its associated covector (field). A
tetrad $u,x,y,z \in T_{m}M$ is called {\em orthonormal} if and only
if the only non-vanishing inner products between tetrad members are
$-g(u,u)=g(x,x)=g(y,y)=g(z,z)=1$ and a tetrad $l,n,x,y\in T_{m}M$ is
called {\em null} if and only if the only non-vanishing inner
products between tetrad members are $g(l,n)=g(x,x)=g(y,y)=1$. In
this case, $l$ and $n$ are null vectors. A space-time will be called
{\em non-flat} if $Riem$ does not vanish over any non-empty open
subset of $M$. This is a physical requirement sometimes imposed to
prevent gravitational shielding in Einstein's theory. Round and
square brackets around indices denote the usual symmetrisation and
skew-symmetrisation of those indices. Let $m\in M$ and let
$\Omega_{m}$ denote the $6$-dimensional vector space of all
skew-symmetric tensors at $m$, irrespective of tensor type (that is,
of the positions of the indices) because of the isomorphisms
resulting from the metric $g(m)$. Members of any of these vector spaces
are, somewhat loosely, called \emph{bivectors}. If $F\in \Omega_{m}$
with components $F^{ab}$ then since $F^{ab}$ is skew-symmetric, its
matrix rank is even. If this rank is two, $F$ is called
\emph{simple} and may be written as $F^{ab}=2p^{[a}q^{b]}$ for $p,q
\in T_{m}M$. The 2-dimensional subspace (2-space) of $T_{m}M$
spanned by $p$ and $q$ is uniquely determined by $F$ and called the
\emph{blade} of $F$. This bivector (or its blade) will sometimes be
denoted by $p \wedge q$. A simple bivector $F$ at $m$ is then called
{\em timelike}, {\em spacelike} or {\em null} according as its blade
is, respectively, a timelike, spacelike or null 2-space at $m$. If
$F\in\Lambda_mM$ has rank four it is called {\em non-simple} and may
be written as $F=G+H$ where $G$ and $H$ are simple bivectors with
$G$ timelike and $H$ spacelike and where the blades of $G$ and $H$
are uniquely determined by $F$ and are orthogonal complements of
each other. The union of these blades spans $T_{m}M$. They will be
collectively called the {\em canonical pair of blades} of $F$. The
symbol $*$ will denote the usual duality operator; thus $G$ and $H$
above are duals of each other (up to scalings).

A remark is added here about the notation in this paper. Both a
coordinate and an index-free notation will be used. However, since
many of the calculations involve tensor contractions and a
coordinate approach is generally much shorter for these, the former
will naturally predominate here.

\section{Curvature structure of space-times}

Because of the algebraic symmetries of $Riem$, one may introduce the
{\em curvature map} $f$ from the vector space of bivectors to the
space of type (1,1) (skew-self adjoint with respect to $g(m)$) tensors
at $m$ (and with the latter space identified with the space of
bivectors at $m$ and referred to as such) by
\begin{equation}\label{1}
f: F^{ab}\rightarrow R^{a}_{\  bcd}F^{cd}
\end{equation}
The rank of the linear map $f$ is referred to as the {\em curvature
rank (of $Riem)$} at $m$ \cite{10}. It is clear that if $F$ is a bivector
which is defined and smooth on some open subset of $M$ then so also is $f(F)$
where $f(F)(m)=f(F(m))$. Let $B_m$ denote the range space
of $f$ at $m$ (suitably interpreted according to the agreed
identification of bivectors) so that $\dim B_m$ equals the curvature
rank at $m$ and is $\leq6$. This leads to a convenient algebraic
classification of $Riem$ at $m$ into five mutually exclusive and
disjoint {\em curvature classes} (for further details, see
\cite{10}).

\begin{enumerate}[{\bf {Class}  A}]
\item This covers all possibilities not covered by
classes $\mathbf{B}$, $\mathbf{C}$, $\mathbf{D}$ and $\mathbf{O}$
below. For this class, the curvature rank at $m$ is 2, 3, 4, 5 or 6.

\item This occurs when $\dim B_m=2$ and when $B_m$ is
spanned by a timelike-spacelike pair of simple bivectors with
orthogonal blades (chosen so that one is the dual of the other). In
this case, one can choose a null tetrad $l,n,x,y\in T_mM$ such that
these bivectors are $F=l\wedge n$ and $\overset{*}{F}=x\wedge y$ so
that $F$ is timelike and $\overset{*}{F}$ is spacelike and then
(using the algebraic identity $R_{a[bcd]}=0$ to remove cross terms)
one has, at $m$,
\begin{equation}\label{2}
        R_{abcd}=\alpha F_{ab}F_{cd}+\beta \overset{*}{F}_{ab}\overset{*}{F}_{cd}
\end{equation}
for $\alpha,\beta\in\mathbb{R}$, $\alpha\neq0\neq\beta$.

\item In this case $\dim B_m=2$ or $3$ and $B_m$ may
be spanned by independent simple bivectors $F$ and $G$ (or $F$, $G$
and $H$) with the property that there exists $0\neq r\in T_mM$ such
that $r$ lies in the blades of $\overset{*}{F}$ and $\overset{*}{G}$
(or $\overset{*}{F}$, $\overset{*}{G}$ and $\overset{*}{H}$). Thus
$F_{ab}r^b=G_{ab}r^b(=H_{ab}r^b)=0$ and $r$ is then unique up to a
multiplicative non-zero real number.

\item In this case $\dim B_m=1$. If $B_m$ is spanned
by the bivector $F$ then, at $m$,
\begin{equation}\label{3}
        R_{abcd}=\alpha F_{ab}F_{cd}
\end{equation}
for $0\neq\alpha\in\mathbb{R}$ and $R_{a[bcd]}=0$ implies that $F_{a[b}F_{cd]}=0$ from which it may be checked that $F$ is necessarily simple.

\setcounter{enumi}{14}
\item In this case $Riem$ vanishes at $m$.
\end{enumerate}

This classification is, of course, pointwise and may vary over $M$.
A space-time $(M,g)$ (or some subset of it) which has the same
curvature class at each point of $M$ is said to be \emph{of that
class}. The subset of $M$ consisting of points at which the
curvature class is $\mathbf{A}$ is an open subset of $M$ (\cite{10},
p 393) and the analogous subset arising from the class $\mathbf{O}$
is closed (and has empty interior in the manifold topology of $M$ if
$(M,g)$ is non-flat). It is important in what is to follow that
\emph{the equation $R_{abcd}k^d=0$ at $m$ has no non-trivial
solutions for $k\in T_{m}M$ if the curvature class at $m$ is
$\mathbf{A}$ or $\mathbf{B}$, a unique independent solution (the
vector $r$ above) if the curvature class at $m$ is $\mathbf{C}$ and
two independent solutions if the curvature class at $m$ is
$\mathbf{D}$ (and which span the blade of $\overset{*}{F}$ in}
(\ref{3})). \emph{If $\dim B_m\geq4$ the curvature class at $m$ is
necessarily $\mathbf{A}$}.

The following result will be useful in what is to follow and the
details and proof can be found in \cite{10}.

\begin{Theorem}\label{Theorem1}
Let $(M,g)$ be a space-time, let $m\in M$ and let $h$ be a non-zero second
order, symmetric, type (0,2) (not necessarily non-degenerate) tensor
at $m$ satisfying $h_{ae}R^e_{\ bcd}+h_{be}R^e_{\ acd}=0$.
\begin{enumerate}[(i)]
\item If the curvature class of $(M,g)$ at $m$ is
$\mathbf{D}$ and $u,v\in T_mM$ span the 2-space at $m$ orthogonal to the blade of
$F$ in (\ref{3}) (that is $u\wedge v$ is the blade of
$\overset{*}{F}$), there exists $\phi,\mu,\nu,\lambda\in\mathbb{R}$
such that, at $m$,
    \begin{equation}\label{4}
        h_{ab}=\phi g_{ab}+\mu u_au_b+\nu v_av_b+\lambda(u_av_b+v_au_b)
    \end{equation}
\item If the curvature class of $(M,g)$ at $m$ is
$\mathbf{C}$ there exists $r\in T_mM$ (the vector appearing in the
above definition of class $\mathbf{C}$) and
$\phi,\lambda\in\mathbb{R}$ such that, at $m$,
    \begin{equation}\label{5}
        h_{ab}=\phi g_{ab}+\lambda r_ar_b
    \end{equation}
\item If the curvature class of $(M,g)$ at $m$ is
$\mathbf{B}$ there exists a null tetrad $l,n,x,y$ (that appearing in
the above definition of class $\mathbf{B}$) and
$\phi,\lambda\in\mathbb{R}$ such that, at $m$ (and making use of the
associated completeness relation),
    \begin{equation}\label{6}
        h_{ab}=\phi g_{ab}+\lambda(l_an_b+n_al_b)=(\phi+\lambda)g_{ab}-\lambda(x_ax_b+y_ay_b)
    \end{equation}
\item If the curvature class of $(M,g)$ at $m$ is
$\mathbf{A}$ there exists $\phi\in\mathbb{R}$ such that, at $m$,
    \begin{equation}\label{7}
        h_{ab}=\phi g_{ab}
    \end{equation}
    \end{enumerate}
\end{Theorem}

The proof of theorem 1 is based on the fact that the members of the range $B_m$
of the map $f$ in (\ref{1}) are, as type $(1,1)$ tensors, skew-self
adjoint with respect to $g$, and also skew-self adjoint with
respect to $h$. Thus each $F\in B_m$ satisfies
\begin{equation}\label{8}
g_{ac}F^c_{\ b}+g_{bc}F^c_{\ a}=0 \hspace{1cm} h_{ac}F^c_{\
b}+h_{bc}F^c_{\ a}=0
\end{equation}
It is a consequence of (\ref{8}) that the blade of $F$ (if $F$ is
simple) and each of the canonical pair of blades of $F$ (if $F$ is
non-simple) are eigenspaces of $h$ with respect to $g$, that is, for
$F$ simple, any $k\in T_{m}M$ in the blade of $F$ satisfies
$h_{ab}k^b=\omega g_{ab}k^b$ where the eigenvalue
$\omega\in\mathbb{R}$ is independent of $k$, and similarly for each
of the canonical blades if $F$ is non-simple (but with possibly
different eigenvalues for these blades) \cite{10}.

[It is remarked that if $\tilde{g}$ is another metric on $M$ whose curvature tensor
$\tilde{Riem}$ equals the curvature tensor $Riem$ of $g$ \emph{everywhere
on $M$} then the conditions of this theorem are satisfied for
$h=\tilde{g}(m)$ at each $m\in M$ and so the conclusions also hold except
that now one must add the restriction $\phi\neq 0$ in each case to
preserve the non-degeneracy of $\tilde{g}$ at $m$ and maybe some
restrictions on $\phi,\mu,\nu$ and $\lambda$ if the signature of
$\tilde{g}$ is prescribed. If $(M,g)$ is of curvature class $\mathbf{A}$, (\ref{7})
gives $\tilde{g}=\phi g$ for some (smooth) function $\phi$ on $M$ and the Bianchi identity may be used to show that
$\phi$ is constant on $M$ \cite{10}. Further, if the condition on $g$ and $\tilde{g}$ of having equal curvatures
is weakened to $\tilde{g}_{ab;[cd]}=0$ on $M$, where a semi-colon denotes a $\nabla$-covariant derivative, the Ricci identity for $\tilde{g}$ shows that the condition of theorem 1 is satisfied for $\tilde{g}=h$ on $M$. In particular, from part (iv) above, if $(M,g)$
is of curvature class $\mathbf{A}$ the condition that $g$ and $\tilde{g}$ are conformally related is equivalent to $\tilde{g}_{ab;[cd]}=0$.]

\section{Holonomy theory}
 Let $m\in M$ and for $1\leq k\leq\infty$ let $C_k(m)$
denote the set of all piecewise $C^k$ closed curves starting and
ending at $m$. If $c\in C_k(m)$ let $\tau_c$ denote the vector space
isomorphism of $T_mM$ obtained by parallel transporting, using
$\nabla$, each member of $T_mM$ along $c$. Using a standard notation
associated with curves one defines, for curves $c,c_0,c_1,c_2\in
C_k(m)$, with $c_0$ denoting a constant curve at $m$, the {\em
identity map} $\tau_{c_0}$ on $T_mM$, the {\em inverse}
$\tau^{-1}_c\equiv\tau_{c^{-1}}$ and {\em product}
$\tau_{c_1}\cdot\tau_{c_2}\equiv\tau_{c_1\cdot c_2}$ to put a group
structure on $C_k(m)$, making it a subgroup of $G\equiv
GL(T_mM)(=GL(4,\mathbb{R}))$, called the {\em $k-$holonomy group} of
$M$ at $m$ and denoted by $\Phi_k(m)$. In fact, since $M$ is
connected and also a manifold, it is also path connected and, as a
consequence, it is easily checked that, up to an isomorphism,
$\Phi_k(m)$ is independent of $m$. Less obvious is the fact that
$\Phi_k(m)$ is independent of $k$ ($1\leq k\leq\infty$) and thus one
arrives at the {\em holonomy group} $\Phi$ (of $\nabla$) on $M$. It can now be proved that $\Phi$ is a Lie group which is connected if $M$ is simply connected. The Lie algebra $\phi$ of $\Phi$ is
called the {\em holonomy algebra}.

If the above operations are repeated this time using only curves in $C_{k}(m)$ homotopic to zero one similarly achieves the \emph{restricted holonomy group} $\Phi^{0}$ of $\nabla$ on $M$ and which can be shown to be the identity component
of $\Phi$ and is equal to $\Phi$ if $M$ is simply connected.
The connection $\nabla$ can be shown to be {\em flat} (that is, $Riem$ vanishes on $M$) if and
only if $\phi$ is trivial. Since, for a space-time $(M,g)$, $\nabla$
is compatible with the metric $g$, that is, $\nabla g=0$, each map
$\tau_c$ on $T_mM$ preserves inner products with respect to $g(m)$.
It can then be shown that $\Phi$ is (isomorphic to) a Lie subgroup
of the {\em Lorentz group} $\mathcal{L}$. Thus the holonomy algebra
$\phi$ can be identified with a subalgebra of the Lie algebra $L$ of
$\mathcal{L}$, the {\em Lorentz algebra}. The one-to-one
correspondence between the subalgebras of $L$ and the
\emph{connected} Lie subgroups of $\mathcal{L}_0$ shows that if $M$
is simply connected the (connected) Lie group $\Phi$ is determined
by the subalgebra of $L$ associated with $\phi$. Further details may
be found in \cite{11} and a summary in \cite{10}.

The subalgebra structure of $L$ is well-known and can be
conveniently represented by informally identifying it with certain
Lie algebras of bivectors in a well-defined way. The binary
operation on $L$ is then that of matrix commutation. Such a
representation of $L$ is well-known and has been classified into
fifteen convenient types \cite{12} (for details of the possible
holonomy types most relevant for the physics of general relativity
see \cite{13,10}). This representation is given in the first three
columns of table 1 using either a null tetrad $l,n,x,y$ or an
orthonormal tetrad $u,x,y,z$ to describe a basis for each
subalgebra. It is noted that in types $R_5$ and $R_{12}$,
$0\neq\omega\in\mathbb{R}$ (and, in fact, $R_5$ cannot occur as the
holonomy algebra for a space-time, but each of the others can - see,
e.g. \cite{10,14}). There is also a type $R_1$, when $\phi$ is
trivial and $(M,g)$ is flat, but this trivial type is omitted. Type
$R_{15}$ is the ``general'' type, when $\phi=L$.
\begin{table}\caption{Holonomy algebras}

\begin{tabular}{|c|c|c|c|c|c|}
\hline
     &        &        &            & Recurrent & Constant  \cr
Type &   Dim. &  Basis & Curvature  & vector    & vector  \cr
     &        &        & Class      & fields    & fields \cr
\hline
 $R_{2}$ & 1 & $l\wedge n$ & $\mathbf{D}$ or $\mathbf{O}$ &$\{l\}, \{n\}$ & $<x, y>$ \\
 $R_{3}$ & 1 & $l\wedge x$ & $\mathbf{D}$ or $\mathbf{O}$ & - & $<l, y>$ \\
 $R_{4}$ & 1 & $x\wedge y$ & $\mathbf{D}$ or $\mathbf{O}$ &- & $<l, n>$ \\
 $R_{5}$ & 1 & $l\wedge n+\omega (x\wedge y$) & - &- & - \\
 $R_{6}$ & 2 & $l\wedge n, l\wedge x$ & $\mathbf{C},\mathbf{D}$ or $\mathbf{O}$ & $\{l\}$ & $<y>$ \\
 $R_{7}$ & 2 & $l\wedge n, x\wedge y$ & $\mathbf{B},\mathbf{D}$ or $\mathbf{O}$ & $\{l\}, \{n\}$ & - \\
 $R_{8}$ & 2 & $l\wedge x, l\wedge y$ & $\mathbf{C},\mathbf{D}$ or $\mathbf{O}$ & - & $<l>$ \\
 $R_{9}$ & 3 & $l\wedge n, l\wedge x, l\wedge y$ & $\mathbf{A},\mathbf{C},\mathbf{D}$ or $\mathbf{O}$ & $\{l\}$ & - \\
 $R_{10}$ & 3 & $l\wedge n, l\wedge x, n\wedge x$ & $\mathbf{C},\mathbf{D}$ or $\mathbf{O}$ & - & $<y>$ \\
 $R_{11}$ & 3 & $l\wedge x, l\wedge y, x\wedge y$ & $\mathbf{C},\mathbf{D}$ or $\mathbf{O}$ & - & $<l>$ \\
 $R_{12}$ & 3 & $l\wedge x, l\wedge y, l\wedge n+\omega(x\wedge y)$ & $\mathbf{A},\mathbf{C},\mathbf{D}$ or $\mathbf{O}$ & $\{l\}$ & - \\
 $R_{13}$ & 3 & $x\wedge y, y\wedge z, x\wedge z$ & $\mathbf{C},\mathbf{D}$ or $\mathbf{O}$ & - & $<u>$ \\
 $R_{14}$ & 4 & $l\wedge n, l\wedge x, l\wedge y, x\wedge y$ & any & $\{l\}$ & - \\
 $R_{15}$ & 6 & $L$ & any & - & -  \\
 \hline
\end{tabular}
\end{table}

If $M$ is simply connected, $\Phi$ is connected and is then uniquely
determined by the subalgebra of $L$ associated with $\phi$. In any
case, $\Phi$ will be referred to according to its Lie algebra label
as in table 1. The simple connectedness condition will not
necessarily be imposed on $M$; it is often sufficient to have this
condition locally and a simply connected, connected chart domain $V$ is
always available. Then the holonomy algebra of the restricted space-time $(V,g)$
is a subalgebra of $\phi$. However, if $M$ is simply connected, the
(nowhere-zero) covariantly constant and recurrent vector fields
(indicated in table 1 by enclosing a basis for the vector space of
covariantly constant vector fields on $M$ for each holonomy type
inside $<\,>$ brackets in the final column and by listing the
independent {\em properly recurrent} vector fields for each holonomy
type in $\{\ \}$ brackets in the second from last column) can be
taken as globally defined on $M$. If $M$ is not simply connected
each $m\in M$ admits a simply connected and connected open
neighbourhood $V$ on which these (nowhere-zero) covariantly constant
or recurrent vector fields are defined but the holonomy type of the restricted space-time
$(V,g)$ may differ from that of $(M,g)$. [A global nowhere-zero
vector field $X$ on a connected, open subset $V$ of $M$ is called
{\em recurrent} on $V$ if it satisfies $\nabla X=X\otimes w$ for
some global, smooth covector field $w$ on $V$ (the {\em recurrence
1-form}). In this case, the \emph{sign} (including zero) of $g(X,X)$
is constant on $V$ and clearly every non-trivial covariantly
constant vector field on $V$ is recurrent on $V$. Then $X$ is called
{\em properly recurrent} on $V$ if it is recurrent on $V$ and is
such that no function $\alpha:V\rightarrow\mathbb{R}$ exists such
that $\alpha$ is nowhere zero on $V$ and $\alpha X$ is covariantly
constant on $V$. In fact, \emph{any non-null} recurrent vector field on
$(M,g)$ with the latter of arbitrary signature and dimension, or \emph{any}
recurrent vector field on a manifold of arbitrary dimension and with
positive definite metric can be globally scaled to be nowhere zero
and covariantly constant because if $Y$ is any such vector field and
$\nabla Y=Y\otimes r$ for some 1-form $r$, $\alpha Y$ is covariantly constant, where
$\alpha=|(g(Y,Y)|^{-\frac{1}{2}}$. Thus properly recurrent vector
fields on $V$ are null everywhere on $V$.] A study of the vector
space of \emph{global, covariantly constant, type $(0,2)$ symmetric
tensor fields} (or local such tensor fields on a fixed open subset of
$M$) for each holonomy type can be obtained from the above theory
and full details can be found in \cite{10,15}. It will be of
importance later. In particular, for the holonomy types $R_{9}$,
$R_{12}$, $R_{14}$ and $R_{15}$, the only such global tensors are
constant multiples of the metric tensor $g$.

A recurrent vector field spans a 1-dimensional distribution on its domain of definition
which is preserved by parallel transport. There is an important
generalisation of this concept. Let $m\in M$ and $W$ a non-trivial
proper subspace of $T_mM$. Suppose $\tau_c(W)=W$ for each $\tau_c$
arising from $c\in C_k(m)$ at $m$. Then $W$ is {\em holonomy
invariant} and gives rise in an obvious way, by parallel transport,
to a smooth distribution on $M$ which is, in fact, integrable
\cite{11}. Clearly, if $W\subset T_mM$ is holonomy invariant then so
is the orthogonal complement, $W^{\perp}$, of $W$. If such a $W$
exists the holonomy group $\Phi$ of $M$ is called {\em reducible}
(otherwise, {\em irreducible}). More details can be found in
\cite{11,16} (see also \cite{10,17}). Thus, for example, in table 1 the
holonomy type $R_2$ admits, locally, two 1-dimensional, null holonomy
invariant subspaces spanned by $l$ and $n$ and which give rise,
locally, to two null, properly recurrent vector fields and infinitely
many 1-dimensional spacelike holonomy invariant subspaces spanned by
the infinitely many covariantly constant vector fields in $<x,y>$.
For the holonomy type $R_7$, two 1-dimensional null holonomy
invariant subspaces exist locally and which give rise, locally, to
two independent properly recurrent null vector fields as in the
previous case, together with a 2-dimensional spacelike one
orthogonal to each of the null ones. For holonomy types $R_{10}$,
$R_{11}$ and $R_{13}$ one has a local 1-dimensional holonomy
invariant subspace, spanned by a covariantly constant vector field
in each case, together with its (3-dimensional) orthogonal
complement. These holonomy decompositions will be useful in what is
to follow.

It is useful, at this point, to introduce the {\em infinitesimal
holonomy group} $\Phi'_m$, of $(M,g)$ at any $m\in M$. Again using a
semi-colon to denote a $\nabla-$covariant derivative, consider, in
some coordinate neighbourhood of $m$, the following matrices for
$(M,g)$ at $m$
\begin{equation}\label{9}
R^a_{\ bcd}X^cY^d,\ \ \ R^a_{\ bcd;e}X^cY^dZ^e,\ \ \ldots
\end{equation}
for $X,Y,Z,...\in T_mM$. It turns out that the collection (\ref{9})
spans a subalgebra of the holonomy algebra $\phi$ (and hence only a
finite number of independent terms arise in (\ref{9}) \cite{11}).
This algebra is called the {\em infinitesimal holonomy algebra at
$m$} and is denoted by $\phi'_m$. The unique connected Lie subgroup
of $\Phi$ that it gives rise to is the {\em infinitesimal holonomy
group} $\Phi'_{m}$ at $m$. Thus the range space of the map $f$ in
(\ref{1}) is, at each $m\in M$, (isomorphic as a vector space to) a
subspace of $\phi$. This gives a restriction, when $\phi$ is known,
on $f$, the expression for $Riem$ at each $m$ and also on the
curvature class of $Riem$ at $m$. This last restriction is listed in the
fourth column of table 1. In addition, it is remarked here
for later convenience that if $(M,g)$ has holonomy type $R_{14}$ it cannot be of curvature class $\mathbf{B}$ everywhere
(see the comments before theorem 4). A useful relationship between the various
algebras $\phi'_m$, the algebra $\phi$ and the curvature class
(through the range space $B_{m}$) at each $m\in M$ is provided by
the Ambrose-Singer theorem \cite{18} (see also \cite{11}). This
theorem says, roughly speaking, that the holonomy algebra of $(M,g)$
can be constructed by first choosing (any) $m\in M$ and, for each
$m'\in M$ and each piecewise $C^{1}$ curve $c$ from $m'$ to $m$, computing the range
space of $f$ at $m'$ and parallely transporting each member of it to
$m$ along $c$. If this is done for each such $m'$ and $c$, the
collection of bivectors accumulated at $m$ spans the holonomy
algebra of $(M,g)$.

\section{Projective structure}

In general relativity the study of projective structure is motivated
by the Newton-Einstein principle of equivalence. Consider the
following question; for a space-time $(M,g)$ with Levi-Civita
connection $\nabla$, if one knows the paths of all the {\em
unparameterised} geodesics (that is, only the {\em geodesic paths}
in $M$ without regard to their parameters) how tightly are $\nabla$ and $g$ determined? Suppose that
$(M,g)$ is a space-time and $g'$ is another metric on $M$ of
arbitrary signature, with respective Levi-Civita connections
$\nabla$ and $\nabla'$, such that the sets of geodesic paths of
$\nabla$ and $\nabla'$ coincide. Then $\nabla$ and $\nabla'$ (or $g$
and $g'$, or $(M,g)$ and $(M,g')$) are said to be {\em projectively related} (on $M$)). [In
fact, it is sufficient that, for each $m\in M$, $\nabla$ and $\nabla'$ share a
non-empty subset of unparametrised $g$-timelike geodesics whose directions at $m$
span a non-empty open subset in the usual
topology on the collection of 1-dimensional subspaces (directions)
of $T_{m}M$ \cite{3}.] Although, in general, a
projectively related pair $\nabla$ and $\nabla'$ may still be
expected to differ, it turns out that in many interesting situations
they are necessarily equal. Intuitively, one might expect a link
between projective relatedness and holonomy type and if
$\nabla=\nabla'$ is the result, $(M,g)$ and $(M,g')$ have the same
holonomy type.
Holonomy theory can then describe precisely the (simple)
relationship between $g$ and $g'$ (and the signatures of $g$ and $g'$ may differ) \cite{15,10}. If $\nabla=\nabla'$,
$g$ and $g'$ will be referred to as \emph{trivially
projectively related} (or \emph{affinely related}). For later convenience in the construction of
examples, the concept of {\em local projective relatedness} will be
required. For a space-time $(M,g)$ let $U$ be a non-empty connected
open subset of $M$ and let $g'$ be some metric defined on $U$. Then
$g$ and $g'$ (or their respective Levi-Civita connections, etc) will be
said to be {\em (locally) projectively related} (on $U$) if (the
restriction of) $g$ on $U$ is projectively related to $g'$ on $U$.
[Generally speaking, this paper is concerned with the situation when
$g'$ is also of Lorentz signature but this will not be assumed.]

If $\nabla$ and $\nabla'$ are projectively related then there exists
a uniquely defined, global, smooth 1-form field $\psi$ on $M$ such
that, in any coordinate domain of $M$, the respective Christoffel
symbols of $\nabla$ and $\nabla'$ satisfy \cite{19,20,21}
\begin{equation}\label{10}
\Gamma'^a_{\ bc}-\Gamma^a_{bc}=\delta^a_{\ b}\psi_c+\delta^a_{\
c}\psi_b
\end{equation}
Conversely, if this condition is satisfied for some global 1-form $\psi$ on $M$,  $\nabla$ and $\nabla'$ are projectively related. It is a consequence of the fact that $\nabla$ and $\nabla'$ are {\em
metric connections} that $\psi$ is a {\em global} gradient on $M$
(see, e.g. \cite{21}). Equation (\ref{10}) can, by using the
identity $\nabla'g'=0$, be written in the equivalent form
\begin{equation}\label{11}
g'_{ab;c}=2g'_{ab}\psi_c+g'_{ac}\psi_b+g'_{bc}\psi_a
\end{equation}
Equation (\ref{10}) reveals a relation
between the type $(1,3)$ curvature tensors $Riem$ and $Riem'$ of
$\nabla$ and $\nabla'$, respectively, given by
\begin{equation}\label{12}
R'^a_{\ bcd}=R^a_{\ bcd}+\delta^a_{\ d}\psi_{bc}-\delta^a_{\
c}\psi_{bd} \hspace{1cm} (\Rightarrow R'_{ab}=R_{ab}-3\psi_{ab})
\end{equation}
where $\psi_{ab}\equiv \psi_{a;b}-\psi_a\psi_b=\psi_{ba}$ and where
$R'_{ab}\equiv R'^{c}{}_{\ acb}$ are the Ricci tensor components of
$\nabla'$.

A particularly important case of such a study arises where the
original pair $(M,g)$ is a space-time which is also an {\em Einstein
space} so that the Ricci and metric tensors are related by
$Ricc=\frac{R}{4}g$. This problem has been discussed in several
places (see the bibliography in \cite{6}). The particular case which
is, perhaps, of most importance in general relativity arises when
the Ricci scalar vanishes and then $(M,g)$ is a \emph{vacuum}
(\emph{Ricci-flat}) space-time and this is discussed in
\cite{1,3,5,8}. It turns out that if $(M,g)$ is a space-time which
is an Einstein space and if $g'$ is another metric on $M$
projectively related to $g$, either $(M,g)$ and $(M,g')$ are each of
constant curvature, or the Levi-Civita connections $\nabla$ and
$\nabla'$ of $g$ and $g'$, respectively, are equal. Since if one of
$(M,g)$ and $(M,g')$ is of constant curvature, so is the other
\cite{21}, then if $(M,g)$ is not of constant curvature, (and so
$\nabla'=\nabla$) an argument from holonomy theory can be used to
show that, \emph{generically}, $(M,g')$ is also an Einstein space
\emph{and} $g'=cg$ ($0\neq c\in\mathbb{R}$). Although examples exist
where each of these conclusions fail, $g'$ always has Lorentz
signature (up to an overall minus sign). If, in addition, $(M,g)$ is
assumed vacuum and the non-flat condition is imposed on it, then
necessarily $\nabla=\nabla'$ and, with one very special case
excluded (the pp-waves!), $g'=cg$ on $M$ ($0\neq c\in\mathbb{R}$)
(and so $(M,g')$ is also vacuum). For this case $g'$ has the same
signature as $g$ (up to an overall minus sign \cite{5}). This result
is important for describing the power of the equivalence principle
in general relativity theory. A similar restrictive result for
space-times of certain holonomy types which are not Einstein spaces will be established in the
next two sections.

With $(M,g)$ given, the object is to find all pairs $(g', \psi)$
satisfying (\ref{11}) where, it is recalled, $\psi$ is a global
gradient, $\psi=d\chi$, for some smooth function
$\chi:M\rightarrow\mathbb{R}$. It is, however, convenient to replace
the pair $(g', \psi)$ by the pair $(a,\lambda)$ where $a$ is a global type
$(0,2)$ symmetric tensor field and $\lambda$ a global 1-form field
on $M$ and which are related to the previous pair by the
Sinjukov transformation \cite{2}

\begin{equation}\label{13}
a_{ab}=e^{2\chi}g'^{cd}g_{ac}g_{bd} \hspace{1cm}
\lambda_a=-e^{2\chi}\psi_bg'^{bc}g_{ac} \ (\Rightarrow
\lambda_a=-a_{ab}\psi^b)
\end{equation}
where a temporary abuse of notation has been used in that $g'^{ab}$ denotes
the contravariant components of $g'$ (and not the tensor $g'_{ab}$
with indices raised using $g$) so that $g'_{ac}g'^{cb}=\delta_a^{\
b}$. Then (\ref{13}) may be inverted to give
\begin{equation}\label{14}
g'^{ab}=e^{-2\chi}a_{cd}g^{ac}g^{bd} \hspace{1cm}
\psi_a=-e^{-2\chi}\lambda_bg^{bc}g'_{ac} \
\end{equation}
Given that $g$ and $g'$ are projectively related metrics on $M$, so
that (\ref{11}) holds for some 1-form $\psi(=d\chi)$, then $a$ and
$\lambda$ as defined in (\ref{13}) can be shown, after a short
calculation, to satisfy Sinjukov's equation
\begin{equation}\label{15}
a_{ab;c}=g_{ac}\lambda_b+g_{bc}\lambda_a
\end{equation}
Now (\ref{13}) implies that $a$ is non-degenerate. It is, in fact,
of the same signature as $g'$, this being easily checked by noting
that at each $m\in M$ the map $T_{m}M \rightarrow T_{m}M$ given by
$k^{a}\rightarrow k'^{a}=e^{\chi} g'^{ab}g_{bc}k^{c}$ is a vector
space isomorphism and an isometry between the (generalised) inner
product spaces $(T_{m}M, a(m))$ and $(T_{m}M, g'(m))$ since
$g'_{ab}k'^{a}k'^{b}=a_{ab}k^{a}k^{b}$. A contraction of (\ref{15})
with $g^{ab}$ then shows that $\lambda$ is a global gradient on $M$
(in fact, of $\tfrac{1}{2}a_{ab}g^{ab}$). It is remarked for future
reference that if the right hand side of (\ref{15}) vanishes at any
$m\in M$ so also does $\lambda(m)$ (as a simple contraction of
(\ref{15}) with $g^{ac}$ shows). In summary, the vanishing of $\lambda$ on $M$ is equivalent to the vanishing of $\psi$ on $M$ and also to the statement that $\nabla=\nabla'$ on $M$.

In practice, to determine which pairs $(g',\nabla')$ are
projectively related to some original pair $(g,\nabla)$ on $M$, it
is often easier to use (\ref{15}) to attempt to find $a$ and
$\lambda$ rather than (\ref{11}) to find $g'$ and $\psi$. This then
requires converting back from $a$ and $\lambda$ to $g'$ and $\psi$. For this purpose the following lemma (the proof of which may be of interest in its own right) is useful. In it, the symmetric non-degenerate type
$(2,0)$ tensor $a^{-1}$ on $M$ is, for each $m\in M$, the
inverse of $a$ ($a_{ac}(a^{-1})^{cb}=\delta_a^{\ b}$). Then raising
and lowering indices on $a$ and $a^{-1}$ with $g$ in the usual way,
one has a global type (0,2) tensor (also labelled $a^{-1}$) with
components $a^{-1}_{ab}=a^{-1cd}g_{ac}g_{bd}$ (and so
$a^{-1}_{ac}a^{cb}=\delta_a^{\ b}$).

\begin{Lemma}\label{Lemma1} (cf \cite{22})

Given a solution pair $(a, \lambda)$ of (\ref{15}) for $(M,g)$ one may associate with it a unique solution pair $(g', \psi)$ of (\ref{11}) by the following procedure. First construct the function $F:M\rightarrow \mathbb{R}$ by $ F=|\det g|/|\det a|$ and then define a type $(0,2)$ metric $g'$ by $g'=Fa^{-1}$. Then define a function $\chi:M\rightarrow \mathbb{R}$ by $\chi=\tfrac{1}{2}\ln F$ and finally define a 1-form $\psi$ by $\psi=d\chi$. The pair $(g', \psi)$ is then a solution to (\ref{11}) and together with the original pair $(a, \lambda)$, satisfies (\ref{13}) and (\ref{14}).
\end{Lemma}

$\mathbf{Proof}$

Define a connection $\tilde{\nabla}$ on $M$ by its Christoffel symbols $\tilde{\Gamma^{a}_{bc}}$ in any coordinate system which are given in terms of the Christoffel symbols of $\nabla$ by $\tilde{\Gamma^{a}_{bc}}=\Gamma^{a}_{bc}+g_{bc}(a^{-1})^{ad}\lambda_{d}$. Then (\ref{15}) implies $\tilde{\nabla} a=0$ and so $\tilde{\nabla}$, being symmetric, is the Levi-Civita connection for the tensor $a$ when the latter is taken as a metric on $M$. Now a standard result gives  $\Gamma^a_{ab}=\tfrac{\partial}{\partial x^b}\left(\tfrac{1}{2}\ln|detg|\right)$ and so, applying this to $\tilde{\nabla}$
\begin{equation}\label{NEW1}
\tfrac{\partial}{\partial x^b}\left(\tfrac{1}{2}\ln|\det a|\right)=\tfrac{\partial}{\partial x^b}\left(\tfrac{1}{2}\ln|\det g|\right)+(a^{-1})_{bd}\lambda^d
\end{equation}
Now define $\psi_b\equiv-(a^{-1})_{bd}\lambda^d$ so that $\psi=d\chi$ is a global gradient, $\chi\equiv\tfrac{1}{2}\ln\left(|\det g|/|\det a|\right)$. Then since $(a^{-1})_{ad}a^{de}=\delta^{d}_{a}$ a covariant differentiation and use of (\ref{15}) gives \cite{22}
\begin{equation}\label{NEW2}
(a^{-1})_{ab;c}=(a^{-1})_{bc}\psi_a+(a^{-1})_{ac}\psi_b
\end{equation}
Then $g'\equiv Fa^{-1}=e^{2\chi}a^{-1}$ and $\psi$ can be shown to satisfy (\ref{11}). It is straightforward to check that (\ref{13}) and (\ref{14}) hold and that $\psi_a=-a^{-1}_{ab}\lambda^b$ (and so $\lambda_a=-a_{ab}\psi^b$). $\square$

It follows that \emph{all projectively related metrics $g'$ together with their
associated 1-forms $\psi$ will be found if all pairs $(a,\lambda)$
can be found satisfying (\ref{15}) and with $a$ non-degenerate}. [In the above definition of $g'(\equiv Fa^{-1})$ it is pointed out that, in \cite{22}, a typographical error arose which resulted in $g'$ being erroneously written as $g'=e^{2\chi}a (=Fa)$.]

Denote the collection of \emph{all} solution pairs $(g', \psi)$ of (\ref{11}) with $g'$ a Lorentz metric on $M$, as described above by $Q'$ and let $P'$ denote the collection of \emph{all} pairs $(a, \lambda)$ where $a$ is a type $(0,2)$ non-degenerate tensor and $\lambda$ a global 1-form on $M$ and which, together, satisfy (\ref{15}). Then $\psi$ and
$\lambda$ are necessarily exact 1-forms on $M$.
The sets $P'$ and $Q'$ contain a certain amount of ``triviality''. For example, if $(g', \psi)$
satisfies (\ref{11}) so also does $(\alpha g', \psi)$ $(0\neq \alpha
\in \mathbb{R})$ and if $(a, \lambda)$ satisfies (\ref{15}) so also
does $(\beta a, \beta \lambda)$ ($0\neq \beta \in \mathbb{R})$.
Thus if one defines a relation
on each of the sets $P'$ and $Q'$ (each denoted by $\approx$ and
readily shown to be an equivalence relation) by $(g'_{1},
\lambda_{1}) \approx (g'_{2}, \lambda_{2})\Leftrightarrow
g'_{2}=\alpha g'_{1}$ and $\lambda_{1}=\lambda_{2}$, $(0\neq \alpha
\in \mathbb{R})$ and $(a_{1}, \lambda_{1}) \approx (a_{2},
\lambda_{2}) \Leftrightarrow a_{2}=\beta a_{1}$ and
$\lambda_{2}=\beta \lambda_{1}$ $(0\neq \beta \in \mathbb{R})$ then
(\ref{13}) and (\ref{14}) give rise to a bijective
correspondence between the quotient sets $P'/\approx$ and
$Q'/\approx$. Each metric in an equivalence class
in $Q'$ has the same connection.

Next, if $(a, \lambda) \in P'$ then so is $(a+h, \lambda)$ where $h$
is a type $(0,2)$ covariantly constant, symmetric tensor on $M$
satisfying the condition that $a+h$ is non-degenerate (and the
choice $h=\gamma g$ for $\gamma$ in some open interval of
$\mathbb{R}$ containing $0$ satisfies this condition). If $(a_{1},
\lambda_{1}),\,(a_{2},\lambda_{2}) \in P'$ write $(a_{1},
\lambda_{1}) \sim (a_{2},\lambda_{2}) \Leftrightarrow a_{2}=a_{1}+h$ for some $h$ as above
(equivalently, $\lambda_{2}=\lambda_{1}$ from (\ref{15}) since now
$\nabla a_{1}=\nabla a_{2}$ (see the remark after (\ref{15}))). Then
$\sim$ is also an equivalence relation on $P'$ and the equivalence
class containing $(a, \lambda)$ is denoted by $[a, \lambda]$. The
associated quotient set is $P'/\sim$.

Now let $P$ denote the set of all solutions to (\ref{15}) where $a$
is restricted only by being a type $(0,2)$ symmetric tensor on $M$,
no requirement of non-degeneracy being imposed. The equivalence
relation $\sim$ described in the previous paragraph  naturally
extends to $P$ to give the quotient set $P/\sim$. Now the set $P$
has a natural structure of a real vector space according to the
operations $(a, \lambda)+(b, \mu)=(a+b, \lambda+\mu)$ and $\gamma(a,
\lambda)=(\gamma a, \gamma \lambda)$ $(\gamma\in \mathbb{R})$. The
zero vector is $(0,0)$ and $-(a, \lambda)= (-a, -\lambda)$. The
subset $T=\{(h, 0)\in P: \nabla h=0\}$ of $P$ is a subspace (the
\emph{trivial subspace}) of $P$. It then follows that $P/\sim$ is
the quotient space of $P$ with respect to $T$ and admits a real
vector space structure according to the operations $[a, \lambda]+[b,
\mu]=[a+b, \lambda + \mu]$ and $\gamma[a, \lambda]=[\gamma a, \gamma
\lambda]$ which are easily checked to be well-defined and the zero
of the vector space $P/\sim$ is $[h, 0](=[g, 0]=[0, 0])$. Next,
consider the map $\sigma:P\rightarrow P/\sim$ given by $\sigma:(a,
\lambda)\rightarrow [a, \lambda]$. Clearly $\sigma$ is linear and
its kernel, $\ker  (\sigma)$, $(=T)$ is (isomorphic to) the vector space of
all type $(0,2)$, symmetric, covariantly constant tensors on $M$.
Thus, $\dim\ker (\sigma)\geq 1$ and, since $\sigma$ is clearly onto,
$\dim P=\dim(P/\sim)+\dim(\ker(\sigma))$. The necessity of dropping
the non-degeneracy condition on $a$ to achieve the vector space
structure on $P$ is clear. It will be seen later (theorem 2) that $P$ is finite dimensional.


 Now apply the Ricci identity to $a$ and use (\ref{15}) to get
\begin{equation}\label{16}
(a_{ab;cd}-a_{ab;dc}=)a_{ae}R^e_{\ bcd}+a_{be}R^e_{\
acd}=g_{ac}\lambda_{bd}+g_{bc}\lambda_{ad}-g_{ad}\lambda_{bc}-g_{bd}\lambda_{ac}
\end{equation}
where $\lambda_{ab}=\lambda_{a;b}(=\lambda_{ba})$. This leads to the
following lemma which is (mostly) a special case of a more detailed
result in \cite{3,22} and for which a definition is required.
Suppose $m\in M$, that $F\in\Lambda_mM$ and that the curvature
tensor $Riem$ of $(M,g)$ satisfies $R^{ab}_{\ \ cd}F^{cd}=\alpha
F^{ab}$ ($\alpha\in\mathbb{R}$) at $m$, so that $F$ may be regarded as a (real)
eigenvector of the map $f$ in (\ref{1}) (recalling the liberties
taken in identifying bivectors described in section 1). Then $F$ is
called a (real) {\em eigenbivector of $Riem$ at $m$ with eigenvalue
$\alpha$}.

\begin{Lemma}\label{Lemma2}
Let $(M,g)$ and $(M,g')$ be space-times with $g$ and $g'$
projectively  related. Suppose at $m\in M$ that $F\in\Lambda_mM$ is
a (real) eigenbivector of $Riem$ of $(M,g)$ with zero eigenvalue (so
that $F$ is in the kernel, $\ker f$, of $f$ in (\ref{1})). Then the
blade of $F$ (if $F$ is simple) or each of the canonical pair of
blades of $F$ (if $F$ is non-simple) is an eigenspace of the
symmetric tensor $\nabla \lambda$ with respect to $g$ at $m$. (That
is, if $p\wedge q$ is any of these blades ($p,q\in T_mM$) there
exists $\mu\in\mathbb{R}$ such that for any $k\in p\wedge q$,
$\lambda_{ab}k^b=\mu g_{ab}k^b$). In particular, if $\ker f$ is such
that $T_mM$ is forced to be an eigenspace of $\nabla \lambda$ at each
$m'$ in some connected open neighbourhood $U$ of $m$, then each of
the following conditions is satisfied on $U$ for some
$c\in\mathbb{R}$
\begin{equation}\label{17}
(a)\ \lambda_{ab}=cg_{ab}, \hspace{1cm} (b)\ \lambda_d
R^{d}{}_{abc}=0, \hspace{1cm} (c)\
a_{ae}R^{e}{}_{bcd}+a_{be}R^{e}{}_{acd}=0
\end{equation}
If (\ref{17}) holds on $U$ either (i) $\lambda$ vanishes
identically on $U$ or (ii) it does not, in which case any point of
$U$ at which it does vanish is topologically isolated in $U$ (and in $M$) and so the subset
of such points constitutes a closed subset of $U$ with empty
interior. Further, $Riem$ vanishes on some neighbourhood of any such
point.
\end{Lemma}
$\mathbf{Proof}$

The proof can mostly be found in \cite{22} except for the last part.
This follows by noting that, from (\ref{17})(a), if $c=0$, $\lambda$
is covariantly constant on $U$ and either vanishes everywhere or
nowhere on $U$. If $c\neq 0$, however, \emph{the vector field with
components $\lambda^{a}$ is proper homothetic with vanishing
homothetic bivector on $U$}. The result now follows from \cite{10}
(theorem 10.6). $\square$

Lemma \ref{Lemma2} gives important algebraic information about projective
structure and it is remarked that any space-time $(M,g)$ which is either of
any holonomy type except $R_{9}$, $R_{14}$ or $R_{15}$ or is of
curvature class $\mathbf{B}$ $\mathbf{C}$ or $\mathbf{D}$ (at each $m\in M$)
necessarily satisfies the conditions leading to (\ref{17}) on $M$ (see table 1 for the relationship between
the holonomy type and the curvature class). If such is the case, either $\lambda$ vanishes on $M$ ($\Leftrightarrow \nabla=\nabla'$) or $(M,g)$ admits
a covariantly constant or proper homothetic vector field and a standard property of (any) homothetic vector field is that it either vanishes identically on $M$ or it cannot
vanish on any non-empty open subset of $M$.

Next, some differential information will be provided which gives a
uniqueness theorem for solutions of (\ref{15}). Returning to
(\ref{16}) a contraction with $g^{ac}$ gives
\begin{equation}\label {NEW3}
4\lambda_{b;d}=\Psi g_{bd}+a^{ec}R_{ebcd}-a_{be}R^e_{\ d}
\end{equation}
where $\Psi\equiv\lambda^a_{\ ;a}$. Then a covariant differentiation of this last equation and use of (\ref{15}) gives
\begin{equation}\label{NEW4}
4\lambda_{b;df}=g_{bd}\Psi_{,f}+a^{ec}R_{ebcd;f}+\lambda^eR_{ebfd}+\lambda^cR_{fbcd}-a_{be}R^e_{\ d;f}-\lambda_{b}R_{df}-g_{bf}\lambda_{e}R^e_{\ d}
\end{equation}
where a comma denotes a partial derivative. The first term in (\ref{NEW4}) can, using the Ricci identity, be replaced by the terms $4(\lambda_{b;fd}+\lambda^{e}R_{ebdf})$. If this is done and the resulting equation contracted with $g^{bf}$ and use made of the contracted Bianchi identity for $(M,g)$ (which gives $R_{ebcd;f}g^{bf}=R_{ec;d}-R_{ed;c}$) one finds
\begin{equation}\label{NEW5}
3\Psi_{,d}=a^{ec}(R_{ec;d}-2R_{ed;c})-10\lambda_{e}R^e_{\ d}
\end{equation}
This equation, together with (\ref{NEW3}) and (\ref{15})
give a first order system of differential equations for the quantities $\lambda_{a},
\Psi$ and $a_{ab}$. [This result has also been noticed in \cite{6}.] It is remarked here
that the condition that $a$ be non-singular is not used in establishing (\ref{NEW3}) and (\ref{NEW5}) and
so with (\ref{15}), this system applies to the set $P$ of all solution pairs $(a,\lambda)$ of (\ref{15}).
This leads to the following theorem.

\begin{Theorem}\qquad{}
\begin{enumerate}[(i)]
\item If $(a, \lambda), (b,\mu) \in P$ and there exists $m\in M$ such
that $a(m)=b(m)$, $\lambda(m)=\mu(m)$ and
$\Psi(m)=\Psi'(m)$ (where $\Psi'=\mu^{a}_{;a}$) then $a=b$ and
$\lambda=\mu$ on $M$ (that is, $(a, \lambda)=(b, \mu)$ on $M$).

\item The vector space $P$ is finite dimensional and $\dim P\leq 15$.

\item If $(a, \lambda), (b,\mu) \in P$ and if there exists a
non-empty open subset $U\subset M$ such that $b=a+\alpha g$ $(\alpha
\in \mathbf{R}$) on $U$ then $b=a+\alpha g$ and $\lambda=\mu$ on
$M$. In particular, if $a=b$ on $U$ then $a=b$ and $\lambda=\mu$ on
$M$ and so $(a, \lambda)=(b,\mu)$.
\end{enumerate}
\end{Theorem}
$\mathbf{Proof}$

(i) Let $I$ be an open interval of $\mathbf{R}$ and $c:I\rightarrow
M$ a smooth curve in $M$ passing through $m$. Let $m'\in c(I)$ and
choose a chart with coordinate functions $x^{a}$ about $m'$.  If $s$
is a parameter for $c$ and $k^{a}=d/ds(x^{a}\circ c$) is the tangent
to $c$ within the chart domain, a contraction of (\ref{15}), (\ref{NEW3}) and (\ref{NEW5})
with $k^{c}$ leads to first order differential equations for the
quantities $\lambda_{a}$, $\Psi$ and $a_{ab}$. Thus if the
conditions of (i) hold at $m'$ the theory of such equations show
that they will hold in an open neighbourhood in $c(I)$ of $m'$. On the other hand, if the conditions of (i) fail
at $m'$ they will, by continuity, fail on some open neighbourhood in $c(I)$  of
$m'$. Since $c(I)$ is a connected subset of $M$ and
since the conditions (i) are given to hold at $m$, they must hold on
$c(I)$. Since $M$ is a manifold whose topology is connected, it is
also path connected and so any point in $M$ may be joined to $m$ by
such a curve. The result now follows.

(ii) This is immediate since the number of independent choices of
$\lambda_{a}(m), \Psi(m)$ and $a_{ab}(m)$ is $4+1+10=15$.

(iii) If the conditions of (iii) hold it is clear from (\ref{15})
that $\lambda$ and $\mu$ agree on $U$. Hence the conditions of part
(i)  above hold at each point of $U$ for $(a+\alpha g, \lambda)$ and
$(b,\mu)$ and the result follows. $\square$

It is clear that this theorem is true (with (ii) obviously modified)
for $(M,g)$ of arbitrary signature and dimension.

Two remarks may be made regarding these results. First,
if $(M,g)$ is an Einstein space, the above differential systems reduce to \cite{5}
\begin{equation}\label{25}
\lambda_{ab;c}=2g_{ab}r_{c}+g_{ac}r_{b}+g_{bc}r_{a} \quad\qquad
(r_{a}=-\frac{R}{12}\lambda_{a})
\end{equation}
and which is precisely
the condition that $\lambda$ is a \emph{projective} (co)vector
field on $(M,g)$ (and is proper projective if $\lambda$ is not identically zero on $M$) and an \emph{affine} (co)vector field in the vacuum
(Ricci flat) case when the constant $R=0$. Thus $\lambda$ vanishes
on $M$ if it vanishes on any non-empty open subset of $M$ (see, e.g.
\cite{10}). Thus, for Einstein spaces, the first order
system has essentially decoupled since (\ref{25}) no longer involves $a$. In this case one must try to find a solution of (\ref{25}) for $\lambda$ and where, in addition, $\lambda$ is an exact 1-form on $M$ and then, if such a $\lambda$ can be found, to find a corresponding solution of (\ref{15}) for $a$. The details of such spaces are known \cite{5,8}. [It is remarked here that
if $(M,g)$ is a space-time which is an Einstein space whose Weyl tensor does not vanish over
any non-empty open subset of $M$ then it cannot admit a proper (that is, not affine) projective vector field \cite{Hall&Lonie2004}]
Second, if the conditions of lemma \ref{Lemma2}, including (\ref{17}), hold then decoupling again takes place and
the above first order system reduces to the homothetic condition for $\lambda$ on $U$
with zero homothetic bivector.
Then $\lambda$ is a homothetic (co)vector field on
$M$ and vanishes on $M$ if it vanishes on any non-empty open subset
of $M$. [The decoupling referred to here is perhaps more obvious if one writes the first order system above in terms of (the 24 independent components of) $\lambda$, $\nabla \lambda$ and $a$ and which can be achieved in a similar manner to that for the original first order system.] Thus, if $(M,g)$ is an Einstein space or satisfies (\ref{17}) in lemma \ref{Lemma2} it can only admit a non-trivially projectively related partner $(M,g')$ if it admits some kind of non-trivial symmetry vector field.

\section{Projective Structure and holonomy; general results}

In this section a summary of some known results and proofs of some
new ones is given and which relate the holonomy type of a space-time
$(M,g)$ to the collection of space-times $(M,g')$ where $g'$ is
(locally or globally) projectively related to $g$. Since the holonomy type of the original
space-time may not be known, these results will also be given in
terms of the curvature class and also in terms of the rank of the
map $f$ (that is, the dimension of the spaces $B_{m}$) in (\ref{1}). This can be
done quite efficiently using the holonomy theory
described in section 3 and allows for easier access to them directly
through the curvature tensor. First the following technical lemma is
required and which was given in \cite{22}.

\begin{Lemma}\label{Lemma3}\qquad{}
\begin{enumerate}[(i)]
\item Let $X$ be a topological space and let $A$ and $B$ be disjoint
subsets of $X$ such that $A$ and $A\cup B$ are open in $X$ and
$A\cup B$ is dense in $X$. Suppose $B=B_{1}\cup B_{2}$, with $B_{1}$
and $B_{2}$ disjoint, and $\mathrm{int}  B\subset
\mathrm{int}  B_{1}\cup B_{2}$ where $\mathrm{int}  $ denotes the
interior operator in $X$. Then $X$ may be
\emph{disjointly} decomposed as $X=A\cup \mathrm{int}  B_{1}\cup
\mathrm{int}  B_{2}\cup J$ where $J$ is the closed subset of $X$
defined by the disjointness of the decomposition and $A\cup
\mathrm{int}  B_{1}\cup \mathrm{int}  B_{2}$ is open and dense in $X$
(that is, $\mathrm{int}   J={\emptyset})$.

\item Let $X$ be topological space and let $A_{1}$, ...,$A_{n}$ be
disjoint subsets of $X$ such that, for
$k=1,...n$, $\cup_{i=1}^{i=k}{A_{i}}$  are open subsets of $X$ and such that
$\cup_{i=1}^{i=n}{A_{i}}$ is (open and) dense in $X$. Then $X$ may
be \emph{disjointly} decomposed as $X=A_{1}\cup \mathrm{int}  
A_{2}\cup...\cup \mathrm{int}  A_{n}\cup K$ where $K$ is the closed
subset of $X$ defined by the disjointness of the decomposition and
$A_{1}\cup \mathrm{int}   A_{2}\cup...\cup \mathrm{int}  A_{n}$ is open
and dense in $X$ (that is, $\mathrm{int}  K={\emptyset}$).
\end{enumerate}
\end{Lemma}

\begin{Theorem}
Let $(M,g)$ and $(M,g')$ be space-times with $\nabla$ and $\nabla'$
projectively related.
\begin{enumerate}[(i)]
\item Suppose $(M,g)$ is of holonomy type  $R_{2}$, $R_{3}$,  $R_{4}$,
$R_{6}$, $R_{7}$, $R_{8}$ or $R_{12}$. Then $\nabla=\nabla'$ on $M$.

\item Suppose $(M,g)$ has holonomy type $R_{10}$, $R_{11}$ or
$R_{13}$ and that there exists $m\in M$ at which the curvature rank
is 2 or 3 (equivalently, at which the curvature class is
$\mathbf{C}$). Then $\nabla=\nabla'$ on $M$.

\item If $M$ admits a non-empty, connected open subset $U$ such that
the only solution of (\ref{15}) on $U$ is $\lambda=0$ and $a=\alpha
g$ $(0\neq \alpha \in \mathbf{R}$) then
the only solution of (\ref{15}) on $M$ is $\lambda =0$ and $a=\alpha
g$ and so $\nabla=\nabla'$. In particular, if $U$ may be chosen so that the holonomy type
of $(U,g)$ is either $R_{9}$ \emph{and} the curvature rank is 1 or 2
at each point of an open dense subset of $U$ (equivalently, $Riem$
is of curvature class $\mathbf{C}$ or $\mathbf{D}$ at each point of
this open dense subset), \emph{or}, $R_{14}$ \emph{and} the
curvature class is $\mathbf{B}$ or $\mathbf{C}$ at each point of an
open dense subset of $U$, $\nabla=\nabla'$ on $M$. Also, if $(M,g)$
is of holonomy type $R_{9}$ \emph{and} $Riem$ has curvature rank 1
or 2 at each point of an open dense subset $U$ of $M$ (equivalently,
$Riem$ is of curvature class $\mathbf{C}$ or $\mathbf{D}$ at each
point of $U$), $\nabla=\nabla'$ on $M$ and if $(M,g)$ is of holonomy
type $R_{14}$ \emph{and} $Riem$ has curvature class $\mathbf{B}$ or
$\mathbf{C}$ at each point of an open dense subset $U$ of $M$,
$\nabla=\nabla'$ on $M$.
\end{enumerate}
\end{Theorem}
$\mathbf{Proof}$

(i) The holonomy conditions imposed show that $(M,g)$ is not flat and so
there exists $m\in M$, and hence an open, connected neighbourhood $U$
of $m$, such that $Riem$ is nowhere zero on $U$. It also follows from the remarks after lemma \ref{Lemma2}
that the conditions of this lemma leading to (\ref{17}) hold at each
point of $M$ and so $\lambda$ is a homothetic (co)vector field on
$M$. Now the holonomy algebra of $(U,g)$ is a subalgebra of that of $(M,g)$
and it follows from table 1 that for each holonomy algebra listed in
part (i), its subalgebras are also contained in this list (excluding, of course, the
$R_{5}$ subalgebra of $R_{12}$ which cannot occur as the holonomy algebra of
a space-time). Thus one may proceed with a hierarchical proof,
using theorems  2, 3 and 5 in \cite{22} applied to the restricted space-time
$(U,g)$. It follows from this that, in each case, $\lambda=0$ on $U$.
Since $\lambda$ is homothetic and vanishes on $U$ it vanishes on $M$ and so $\psi=0$ (and $\nabla=\nabla'$) on $M$.

(ii) If there exists $m\in M$ where the curvature rank is 2 or 3
(and since the maximum curvature rank for these holonomy types is 3)
then by one of the many versions of the rank theorem there exists a
connected, open neighbourhood $U$ of $m$ such that the curvature
rank is 2 or 3 (and hence the curvature class is $\mathbf{C}$) at
each point of $U$. One now proceeds as in part (i) to see that
(\ref{17}) holds on $M$ and so, applying theorem 4 of \cite{22} to $(U,g)$,  $\lambda$ vanishes on $U$
and hence on $M$. Thus $\nabla=\nabla'$ on $M$.

(iii) The first part of (iii) is clear from theorem 2(iii) since the pair $(\alpha g,0)$ is a global solution of (\ref{15})
on $M$ and agrees with any other such solution on $U$.

Next suppose that $(U,g)$ is of holonomy type $R_{9}$. Then it is
noted from table 1 that if $Riem$ is of curvature class $\mathbf{C}$
at $m\in U$, it is necessarily of curvature rank 2 at $m$ with the
vector $r$ in the definition of this curvature class either null or
spacelike. Conversely, if $Riem$ is of curvature rank 2 at $m$ it is
necessarily of curvature class $\mathbf{C}$ since all members of
$B_{m}$ are, for this holonomy type, simple with the null vector $l$
in their blade. It is also remarked that if the curvature class is
$\mathbf{D}$ at any $m\in M$, (\ref{3}) holds with $l$ in the blade
of $F$. So let $J$, $C_{nn}$, $C_{n}$, $D_{nn}$ and $D_{n}$ denote,
respectively, the subsets of points of $U$ at which $Riem$ vanishes,
is of (curvature) class $\mathbf{C}$ with $r$ spacelike, is of class
$\mathbf{C}$ with $r$ null, is of class $\mathbf{D}$ with $F$
timelike in (\ref{3}) and is of class $\mathbf{D}$ with $F$ null in
(\ref{3}). Define the sets $C\equiv C_{nn}\cup C_{n}$ and $D\equiv
D_{nn}\cup D_{n}$ so that, by the rank theorem, $C$ and $C\cup D$
are open in $U$ and, in addition, $C\cup D$ is dense in $U$. Then
consider the disjoint decomposition

\begin{equation}\label{26}
U=C\cup \mathrm{int}   D_{nn}\cup \mathrm{int}  D_{n}\cup K
\end{equation}
where the interior operator is here taken in the subspace topology
of $U$ (but, since $U$ is open in $M$, this is the same as the
interior operator in $M$). In this decomposition the closed subset
$K$ (in the topology of $U$) is determined by the disjointness of
the decomposition. Also, $C_{nn}$ is open in $C$ (and hence in $M$)
because if $C_{nn}\neq \emptyset$ and $m\in C_{nn}$, $B_{m}$
contains a simple timelike member, say $F$. So let $G$ be a smooth
bivector field defined on some open neighbourhood of $m$ and such that
$f(G)(m)=F$. Then $f(G)$ is timelike over some neighbourhood $V$ of $m$. Since $C$ is
open in $U$ one may choose $V\subset C$ and so $V\subset C_{nn}$. It
follows that $C_{nn}$ is open in $C$ and hence in $U$. If $\mathrm{int}  D\neq\emptyset$ then for $m\in\mathrm{int}  D$
there is a connected open neighbourhood $V'$ of $m$ contained in
$\mathrm{int}  D$ on which $Riem$ may be written as in (\ref{3}) with $F$
smooth (see \cite{10}). Then a consideration of the smooth function $F^{ab}F_{ab}$
on $V'$ reveals that $F^{ab}F_{ab}(m)<0$ for $m\in D_{nn}$ and
$F^{ab}F_{ab}(m)=0$ for $m\in D_{n}$. Thus $m\in\mathrm{int}  D\Rightarrow
m\in\mathrm{int}  D_{nn}$ \emph{or} $m\in D_{n}$ and so $\mathrm{int}  D\subset$
$\mathrm{int}  D_{nn}\cup D_{n}$. Since $C$ and $C\cup D$ are each open in $U$
and $C\cup D$ is dense in $U$, lemma \ref{Lemma3} (i) shows that the closed (in $U$) set
$K$ has empty interior (in $U$) and hence that $U\setminus
K$ is open and dense in $U$. It then suffices to show for the first part of the proof that under the relevant
conditions of part (iii), $\lambda$ vanishes on each of $C_{nn}$,
$\mathrm{int}  C_{n}$, $\mathrm{int}  D_{nn}$ and $\mathrm{int}  D_{n}$ (because $\lambda$ then
vanishes on $C_{n}$ and hence on $ C$) to establish that $\lambda$
vanishes on $U\setminus K$ and hence on $U$.

So suppose $C_{nn}\neq \emptyset$ and let $m\in$ $C_{nn}$. Then, by
the $R_{9}$ holonomy type assumption, it can be checked that
there exists an open
neighbourhood $V\subset C_{nn}$ of $m$ and smooth vector fields $l,
n, x$ and $y$ on $V$ such that $l$ is null and recurrent on $V$ (say
$\nabla l=l\otimes p$ for some 1-form field $p$ on $V$), such that
$l, n, x$ and $y$ constitute a null tetrad at each point of $V$ and
such that $l\wedge n$ and $l\wedge x$ span $B_{m'}$ at each $m'\in
M$ (and so $R^{a}{}_{bcd}y^{d}=0$ on $V$). Then, from (\ref{1}),
$l\wedge x$, $l\wedge y$, $x\wedge y$ and $n\wedge y$ span
$\ker (f_{m'})$ at each $m'\in V$ and the conditions and conclusions of
lemma \ref{Lemma2} hold. Thus $\lambda_{a;b}=cg_{ab}$ for some constant $c$ and
$a$ satisfies (\ref{17}(c)). So theorem 1(ii) gives on $V$

\begin{equation}\label{27}
a_{ab}=\phi g_{ab}+\mu y_{a}y_{b}
\end{equation}
where $\phi$ and $\mu$ are functions whose smoothness follows from
that of $a_{ab}y^{a}y^{b}$ and $g^{ab}a_{ab}$. A substitution into
(\ref{15}) (recalling that a comma denotes a partial derivative)
then reveals

\begin{equation}\label{28}
\phi_{,c}g_{ab}+\mu _{,c}y_{a}y_{b}+\mu
(y_{a;c}y_{b}+y_{a}y_{b;c})=g_{ac}\lambda_{b}+g_{bc}\lambda_{a}
\end{equation}
 Contractions of (\ref{28}) with $l^{a}x^{b}$, $l^{a}y^{b}$ and
 $n^{a}x^{b}$ easily show that
 $\lambda_{a}l^{a}=\lambda_{a}n^{a}=\lambda_{a}x^{a}=\lambda_{a}y^{a}=0$
 on $V$ and so $\lambda=0$ on $V$. It follows that $\lambda=0$ on
 $C_{nn}$.

If $\mathrm{int}  C_{n}\neq \emptyset$ let $m\in$ $\mathrm{int}  C_{n}$ and choose an open
neighbourhood $V\subset C_{n}$ of $m$ and smooth vector fields $l,
n, x$ and $y$ on $V$ as before such that $l$ is recurrent, $l\wedge
x$ and $l\wedge y$ span $B_{m'}$ for each $m'\in V$ and
$R^{a}{}_{bcd}l^{d}=0$ on $V$. Then $l\wedge x$, $l\wedge y$,
$l\wedge n$ and $x\wedge y$ span $\ker (f_{m'})$ at each $m'\in V$ and
lemma \ref{Lemma2} again applies. Thus theorem 1(ii) shows that
\begin{equation}\label{29}
a_{ab}=\phi g_{ab}+\mu l_{a}l_{b}
\end{equation}
for functions $\phi$ and $\mu$ on $V$ which are smooth since
$a_{ab}n^{a}n^{b}$ and $g^{ab}a_{ab}$ are. Sinjukov's equation
(\ref{15}) then gives
\begin{equation}\label{30}
\phi _{,c}g_{ab}+2\mu
l_{a}l_{b}p_{c}+\mu_{,c}l_{a}l_{b}=g_{ac}\lambda_{b}+g_{bc}\lambda_{a}
\end{equation}
and contractions with $l^{a}x^{b}$ and $n^{a}y^{b}$ again show that
$\lambda=0$ on $V$ and hence on $\mathrm{int}  C_{n}$. [There is an alternative
proof here that will prove useful in other cases to follow. The
condition $R^{a}{}_{bcd}l^{d}=0$ is easily seen to hold for this case and the recurrence condition
and Ricci identity for $l$ show that $l_{a;[bc]}=0$ and
hence that $p_{[a;b]}=0$ on $V$. Thus the 1-form $p$ is closed and
hence is locally a gradient on $V$. Thus, by reducing $V$ if
necessary, one may write $p_{a}=\alpha _{,a}$ for some function
$\alpha$ on $V$ and then $e^{-\alpha} l^{a}$ is covariantly constant
on $V$. Then the proof for the holonomy type $R_{8}$ given in
\cite{22} shows that $\lambda=0$ on $V$ and hence on  $\mathrm{int}  C_{n}$.]

If $\mathrm{int}  D_{nn}\neq \emptyset$ let $m\in\mathrm{int}  D_{nn}$ and choose an
open neighbourhood $V\subset D_{nn}$ of $m$ and smooth vector fields
$l, n, x$ and $y$ on $V$ as before with $l$ recurrent and with
$l\wedge n$ spanning $B_{m'}$ at each $m'\in V$ and
$R^{a}{}_{bcd}x^{d}=R^{a}{}_{bcd}y^{d}=0$ on $V$. Again lemma \ref{Lemma2}
applies and theorem 1(i) gives
\begin{equation}\label{31}
a_{ab}=\phi g_{ab}+\mu x_{a}x_{b}+\nu y_{a}y_{b}+\rho
(x_{a}y_{b}+y_{a}x_{b})
\end{equation}
for functions $\phi$, $\mu$, $\nu$ and $\rho$ which are then seen to be smooth
since $a_{ab}x^{a}x^{b}$, $a_{ab}y^{a}y^{b}$, $a_{ab}x^{a}y^{b}$ and
$a_{ab}g^{ab}$ are. Sinjukov's equation (\ref{15}) then reveals that
\begin{eqnarray}\label{32}
\phi_{,c}g_{ab}+ \mu_{,c} x_{a}x_{b}+ \mu
(x_{a;c}x_{b}+x_{a}x_{b;c})+\nu _{,c} y_{a}y_{b}+\nu
(y_{a;c}y_{b}+y_{a}y_{b;c})\\\nonumber+\rho
_{,c}(x_{a}y_{b}+y_{a}x_{b})+\rho
(x_{a;c}y_{b}+x_{a}y_{b;c}+y_{a;c}x_{b}+y_{a}x_{b;c})=g_{ac}\lambda_{b}+g_{bc}\lambda_{a}
\end{eqnarray}
Noting that, from the recurrence condition on $l$,
$l^{a}x_{a;b}=l^{a}y_{a;b}=0$ on $V$, a contraction of (\ref{32}) with
$l^{a}$ and a simple rank argument gives $\lambda_{a}l^{a}=0$ and
$\phi_{,a}l_{b}=l_{a}\lambda_{b}$, whilst a contraction with
$n^{a}n^{b}$ gives $\lambda_{a}n^{a}=0$. Hence $\lambda=0$ on $V$
and hence on $\mathrm{int}  D_{nn}$.

If $\mathrm{int}  D_{n}\neq \emptyset$ let $m\in\mathrm{int}  D_{n}$ and choose an open
neighbourhood $V\subset D_{n}$ of $m$ and smooth vector fields $l,
n, x$ and $y$ on $V$, as before, such that $l$ is recurrent and such
that $l\wedge x$ spans $B_{m'}$ at each $m'\in\mathrm{int}  D_{n}$ (and
note that $R^{a}{}_{bcd}l^{d}=R^{a}{}_{bcd}y^{d}=0$ on $V$). The
same procedure as in the previous cases together with contractions
of the final equation with $l^{a}$ and $x^{a}x^{b}$ lead to
$\lambda_{a}l^{a}=0$, $\phi_{,a}l_{b}=l_{a}\lambda_{b}$ and
$\phi_{,a}=2(\lambda_{b}x^{b})x_{a}$ from which it follows that
$\lambda=0$ on $V$ and hence on $\mathrm{int}  D_{n}$.

It follows that
$\lambda=0$ on $U$ and hence, from (\ref{15}) that $\nabla a=0$ on $U$. But since $(U,g)$ is of holonomy type $R_{9}$,
it follows from remarks in section 3 that the only solution of $\nabla a=0$ on $U$ is $a=\alpha g$ for $\alpha\in \mathbb{R}$. The result now follows from the first sentence of part (iii) of this theorem.

If $(U,g)$ has holonomy type $R_{14}$ the restriction that $Riem$
must be of curvature class $\mathbf{B}$ or $\mathbf{C}$ over an open
dense subset $U'$ of $U$ means that the curvature rank is either 2
or 3 over $U'$. The possible spanning bivectors in $B_{m}$ for $m\in
U'$ (table 1) show that, in the notation of table 1, the simple
members of $B_{m}$ cannot contain both $l\wedge n$ and $x\wedge y$
whilst the null members cannot contain either $l\wedge n$ or
$x\wedge y$. [To see this, use the fact that a non-zero bivector
$H\in \Lambda_{m}M$ is simple $\Leftrightarrow
H_{ab}\overset{*}{H^{ab}}=0$ and is null $\Leftrightarrow$
$H_{ab}H^{ab}=H_{ab}\overset{*}{H^{ab}}=0$.] Thus $U$ may be
disjointly decomposed as follows
\begin{equation}\label{33}
U=C^{3}\cup B\cup C^{2}_{nn}\cup C^{2}_{n}\cup J
\end{equation}
where $C^{3}$, $B$, $C^{2}_{nn}$, $C^{2}_{n}$ and $J$ are subsets of
$U$ with $C^{3}$ the subset of points $m'$ at which the curvature class
is $\mathbf{C}$,  $\dim B_{m'}=3$ and $r$ in the definition of
curvature class $\mathbf{C}$ in section 2 is (necessarily) null, $B$
the subset of points $m'\in U$ at which the curvature class is
$\mathbf{B}$, $\dim B_{m'}=2$, $C^{2}_{nn}$ the subset of points
$m'\in U$ at which the curvature class is $\mathbf{C}$,
$\dim B_{m'}=2$ and $r$ spacelike, $C^{2}_{n}$ the subset of points
$m'\in M$ at which the curvature class is $\mathbf{C}$,
$\dim B_{m'}=2$ and $r$ is null and $J$ is closed with empty interior.
(Again all topological statements are made in the subspace topology
of $U$.) The set $C^{3}$ is open in $U$ by the rank theorem as is
$C^{3}\cup B$ (since $B_{m'}$ contains non-simple members if and
only if $m'\in B$ and so, for $m'\in B$ and for some smooth bivector
$H$ defined on some open neighbourhood of $m'$,
$f(H)(m')$ is a non-simple member of $B_{m'}$ and $f(H)$ is non-simple
over some open neighbourhood of $m'$). Clearly, if
$C^{2}\equiv C^{2}_{nn}\cup C^{2}_{n}$, then $C^{3}\cup B\cup C^{2}$
is open in $U$ and use of the function $F^{ab}F_{ab}$ (as in an
argument above) or, if $\mathrm{int}  C^{2}\neq \emptyset$, of the possibility of
choosing a smooth (vector field) solution $k$ of the equation $R^{a}{}_{bcd}k^{d}=0$ on $\mathrm{int}  C^{2}$,
shows that $\mathrm{int}  C^{2}\subset\mathrm{int}  C^{2}_{nn}\cup
C^{2}_{n}$. Thus one has the disjoint decomposition given by
\begin{equation}\label{34}
U=C^{3}\cup B\cup\mathrm{int}  C^{2}_{nn}\cup\mathrm{int}  C^{2}_{n}\cup K
\end{equation}
in which $K$ is closed and has empty interior in $U$, from lemma
\ref{Lemma3}(i). Thus, to show $\lambda$ vanishes on $U$, it is sufficient to show that $\lambda$ vanishes on
each interior in (\ref{34}) (including $C^{3}\cup B$) and hence on the
open dense subset $U\setminus K$ of $U$.

If $C^{3}\neq \emptyset$, let $m\in C^{3}$ and choose an open
neighbourhood $V\subset C^{3}$ of $m$ on which vector fields $l, n,
x$ and $y$ are defined, as before, with $l$ recurrent. It can then
be checked that these vector fields may be chosen such that for each
$m'\in C^{3}$, $B_{m'}$ is spanned by $l\wedge x$ $l\wedge y$ and
$x\wedge y$, evaluated at $m'$. Thus, $R^{a}{}_{bcd}l^{d}=0$ on $V$
and the Ricci identity may be used, as in the (alternative) proof following (\ref{30})
above, to show that $V$ may be reduced, if necessary, so that $l$ is
covariantly constant on $V$. It then follows from the proof for the
holonomy type $R_{11}$ in \cite{22} that $\lambda=0$ on $V$ and
hence on $C^{3}$.

If $B\neq \emptyset$ the proof that $\lambda=0$ on $B$ follows  from
the proof in the holonomy type $R_{7}$ case in \cite{22} (theorem 3).
[It is remarked here that, in \cite{22}, it was stated that a space-time of curvature class $\mathbf{B}$
had holonomy type $R_{7}$. This is true if $\alpha$, $\beta$ and $F$ are smooth in (\ref{2})
but is not proven if they are not. However, theorem 3 in \cite{22} still holds since  $\alpha$, $\beta$ and $F$ can be shown to be smooth over an open dense subset of $M$ on which $\lambda$ then vanishes and so $\lambda=0$ on $M$. In the present case, $(M,g)$ has holonomy type $R_{14}$ and the local existence of the smooth
recurrent vector field $l$ and hence of the associated smooth local null tetrad reveals the smoothness of
$\alpha$, $\beta$ and $F$ on $B$ and the result that $\lambda=0$ on $B$ follows.]
If $\mathrm{int}  C^{2}_{nn}\neq \emptyset$ the proof that $\lambda=0$ on
$\mathrm{int}  C^{2}_{nn}$ is the same as that given above for
holonomy type $R_{9}$ on the subset $C_{nn}$. If $\mathrm{int}  C^{2}_{n}\neq
\emptyset$, the proof that $\lambda=0$ on $\mathrm{int}  C^{2}_{n}$ is as for the
holonomy type $R_{11}$ (curvature class $\mathbf{C}$) given in
\cite{22}. It follows that $\lambda=0$ on $U$ and, since $(U,g)$ is
of holonomy type $R_{14}$, an argument similar to that in the $R_{9}$ case shows that the only solutions
of (\ref{15}) on $U$ are of the form $\lambda=0$ and $a=\alpha g$
for $\alpha\in \mathbb{R}$. The result
now follows.

The proof of the two statements in the final sentence of part (iii) now follows
from the previous results. $\square$

The last theorem can be restated in many other ways involving, say,
curvature class or curvature rank instead of holonomy type. To avoid
too much repetition, the following theorems single out some special
cases.

\begin{Theorem}
Let $(M,g)$ and $(M,g')$ be space-times with $\nabla'$ and $\nabla$
projectively related.
\begin{enumerate}[(i)]
\item Suppose $(M,g)$ is of curvature class $\mathbf{D}$ (equivalently
curvature rank 1) on some open, dense subset $U$ of $M$ and of
holonomy type $R_{2}$, $R_{3}$, $R_{4}$, $R_{6}$, $R_{7}$, $R_{8}$,
$R_{9}$ or $R_{12}$. Then $\nabla'=\nabla$ on $M$.

\item Suppose $(M,g)$ is of curvature class $\mathbf{C}$ on some
open, dense subset $U$ of $M$ and of any permissible holonomy type
except $R_{15}$. Then
$\nabla'=\nabla$ on $M$.

\item Suppose $(M,g)$ is of curvature class $\mathbf{B}$ on some
open, dense subset $U$ of $M$. Then $\nabla'=\nabla$ on $M$.

\item Suppose that $(M,g)$ is such that at each point $m$ of some
non-empty, open, subset $U$ of $M$, the curvature class is
$\mathbf{A}$  and $(\ker f)_{m}$ is such that (\ref{17}) holds at $m$.
Then $\nabla'=\nabla$ on $M$.
\end{enumerate}
\end{Theorem}
$\mathbf{Proof}$

(i) For the holonomy types $R_{2}$, $R_{3}$, $R_{4}$, $R_{6}$,
$R_{7}$, $R_{8}$ and $R_{12}$ the result follows trivially from theorem 3(i). For
the type $R_{9}$ it follows from theorem 3(iii).

(ii) The curvature class $\mathbf{C}$ condition and the consequent fact that
the curvature rank is $\geq 2$ at each $n\in U$ means that the holonomy type is
not $R_{7}$ and, of course, cannot be
$R_{2}$, $R_{3}$ or $R_{4}$. For types $R_{6}$, $R_{8}$, $R_{10}$
$R_{11}$, $R_{12}$ and $R_{13}$, the result follows trivially from theorems 3(i) and 3(ii)
and for types $R_{9}$ and $R_{14}$ it follows from theorem 3(iii).

(iii) This follows by using lemma \ref{Lemma2} and (\ref{17}) to show that $\lambda=0$ on $U$ and hence on $M$.

(iv) This is clear since the curvature class $\mathbf{A}$
restriction on $U$ means that the only solution to (\ref{17}c) is
when $a$ is proportional to $g$ on $U$ \cite{10,15}. On substituting
this into (\ref{15}) and performing some obvious contractions one
finds that $\lambda=0$ on $U$ and the result now follows from the first
part of theorem 3(iii).

\begin{Theorem}\label{Theorem5}
Let $(M,g)$ and $(M,g')$ be space-times with  $\nabla'$ and $\nabla$
projectively related.
\begin{enumerate}[(i)]
\item Suppose $Riem$ has curvature rank 2 at each point of some open
dense subset $U$ of $M$ and $(M,g)$ has any permissible holonomy
type except $R_{15}$. Then $\nabla'=\nabla$ on $M$.

\item Suppose $Riem$ has curvature rank 3 at each point of some open
dense subset $U$ of $M$, that $(\ker f)_{m}$ is such that lemma \ref{Lemma2}
holds and (\ref{17}) is satisfied at each $m\in U$ and that $(M,g)$
has any permissible holonomy type except $R_{15}$. Then
$\nabla'=\nabla$ on $M$.

\end{enumerate}
\end{Theorem}
$\mathbf{Proof}$

(i) Since $Riem$ has curvature rank 2 at each point of $U$ one may
decompose $M$, disjointly, as $M=A\cup B\cup C\cup J$ where $A$
(respectively, $B$, $C$ and $J$) denote the subsets of $M$ at each
point of which the curvature class is $\mathbf{A}$ (respectively
$\mathbf{B}$, $\mathbf{C}$ and $\mathbf{O}$) and $\mathrm{int}  J=\emptyset$. Then $M$ admits the
disjoint decomposition
\begin{equation}\label{35}
M=A\cup\mathrm{int}  B\cup\mathrm{int}  C\cup K
\end{equation}
where $A$ is open in $M$. Since $A\cup B$ and $A\cup B\cup C$ are
open in $M$, with the latter dense in $M$, it follows from lemma \ref{Lemma3}(ii) that the closed set $K$ has
empty interior in $M$ and hence that $M\setminus K$ is open and
dense in $M$. Since  $\dim\ker f=4$ at each $m\in U$ it can be shown that
$B_{m}$ always contains a non-simple member (\cite{10}, p 392) and the conclusions of
lemma \ref{Lemma2} can be checked to apply on $U$. Then theorem 4 parts (iv),
(iii) and (ii)
reveal that $\lambda$ vanishes on $A$, $\mathrm{int}  B$ and $\mathrm{int}  C$,
respectively, if any of these is non-empty. Thus $\lambda=0$ on the
open dense subset $M\setminus K$ of $M$ and hence on $M$.

(ii) Under the conditions stipulated here the curvature class is
either $A$ or $C$ at each point of $U$. One then decomposes M
disjointly in an obvious notation as $M=A\cup C\cup J=A\cup\mathrm{int}  C\cup
K$ where $J$ is closed and has empty interior in $M$ and, since $A$
and $A\cup C$ are open in $M$ with the latter dense in $M$, lemma \ref{Lemma3}(ii) shows that the closed set
$K$ has empty interior in $M$. So $M\setminus K$ is open and dense
in $M$. The result now follows from theorem 4 part (ii) (for $\mathrm{int}  C$)
and part (iv) for $A$. $\square$

In the event that the projective related condition linking $(M,g')$ and $(M,g)$ leads to $\nabla'=\nabla$ an
 argument from holonomy theory can then be used to find the relationship between $g'$ and $g$ \cite{10, 22}.

An inspection of theorems 3, 4 and 5 reveals certain gaps in them
which will now be identified properly. As before, let $(M,g)$ and
$(M,g')$ be space-times, with $\nabla$ and $\nabla'$ projectively
related. First consider the situation when $(M,g)$ has holonomy type
$R_{10}$, $R_{11}$ or $R_{13}$. It is clear from theorem 3(ii) that
a full resolution of these holonomy types requires only a
consideration of the situation when the curvature rank is $\leq 1$ at each point of
$M$, that is, when $M$ is of curvature class
$\mathbf{D}$ or $\mathbf{O}$ at each of its points. This was
completed in \cite{22} where it was shown that, locally or globally,
metrics $g$ and $g'$ exist which are projectively related but where
$\nabla\neq\nabla'$  and, given $g$, all such metrics $g'$ may be
found. The proof in \cite{22} can be improved a little by noting
from the last part of lemma \ref{Lemma2} that, if $U$ is the non-empty open subset
of $M$ on which $Riem$ does not vanish, whether $c$ is zero or not
in (\ref{17}), $\lambda$ is either identically zero on $M$ or nowhere zero on $U$. The problem of
local projective relatedness on $M$ can then be resolved on some neighbourhood of any $m\in U$. If $M$ is non-flat
one may take $U$ open and dense in $M$.
This completes the situation for holonomy types $R_{10}$,
$R_{11}$ and $R_{13}$.

For holonomy type $R_{9}$, and with the non-flat condition assumed
on $(M,g)$ for simplicity, one sees that, at each point of an open dense subset $U$
of $M$, the curvature class of $Riem$ may be $\mathbf{A}$,
$\mathbf{C}$ or $\mathbf{D}$ and theorem 3(iii) shows that
$\nabla=\nabla'$ on $M$ except possibly when there exists $m\in M$
at which the curvature class is $\mathbf{A}$. In this case, since
the subset of points of $M$ at which the curvature class is
$\mathbf{A}$ is open in $M$ (\cite{10}, page 393) one is led to
consider the situation when $M$ admits a non-empty open subset on
which the curvature class is $\mathbf{A}$. This will be considered
in the next section.

For holonomy type $R_{14}$ and with the non-flat condition again assumed
on $(M,g)$ one sees that, at each point of an open dense subset $U$
of $M$, the curvature class may be any of the classes  $\mathbf{A}$,
$\mathbf{B}$, $\mathbf{C}$ or $\mathbf{D}$. Theorem 3(iii) shows
that $\nabla=\nabla'$ on $M$ except possibly when the curvature
class is $\mathbf{A}$ at some $m\in U$ and hence in some open
neighbourhood of $m$ contained in $U$, or the curvature class is
$\mathbf{D}$ at each point of some non-empty open subset $V$ of $U$.
The first of these cases will be dealt with in the next section. For
the second possibility, (\ref{3}) holds on $V$ so let $V'$ be any
(necessarily open) component of $V$ and consider the space-time
$(V',g)$, decomposing it, in a manner done several times above, into
regions where $F$ in (\ref{3}) is spacelike, timelike or null. The
techniques used for holonomy type $R_{6}$ in \cite{22} or $R_{9}$
above show that $\lambda=0$ on $V$ except, possibly, when $F$ is
spacelike over some non-empty open subset of $V$. This latter case
then becomes essentially the same as a subcase of the holonomy type
$R_{11}$ and is dealt with in \cite{22} as mentioned above.

For curvature class $\mathbf{C}$ theorem 4(ii) will be completed in
the next section by exhibiting an example of a space-time $(M,g)$ of
curvature class $\mathbf{C}$ and holonomy type $R_{15}$ which admits
a projectively related metric $g'$ for which $\nabla \neq \nabla'$.
[The well-known set of FRWL cosmological metrics gives examples of
projectively related, but distinct, Levi-Civita connections of
curvature class $\mathbf{A}$ and holonomy type $R_{15}$ \cite{4}]

\section{Projective Structure and Holonomy; Special Cases}

The only space-time holonomy group possibilities that have not been
``essentially'' completely described (that is, neglecting in some
cases the possibility of non-empty flat regions) by the
previous theorems are the types $R_{9}$, $R_{14}$ and $R_{15}$.
These three holonomy types will now be discussed and shown to reveal projective
relationships between Levi-Civita connections which do \emph{not}
result in the equality of these connections.

\subsection{Space-times admitting a normal conformal vector field}

It is convenient, at this point, to introduce a lemma which is a
special case of the following result \cite{23,10}. If $M$ is a
manifold of arbitrary signature $n\geq 2$, $g$ is a metric on $M$ of
arbitrary signature and $M$ admits a global, nowhere zero or null,
hypersurface-orthogonal conformal vector field $X$ then $X$ is
covariantly constant with respect to the Levi-Civita connection of
the conformally related global metric $\tilde{g}=|g(X,X)|^{-1}g$ on
$M$. A similar \emph{local} result holds if $X$ is everywhere null.
[Here, hypersurface-orthogonal means that the covector field
associated with $X$ is locally proportional to a gradient but with
the proportionalty factor not necessarily constant.]

\begin{Lemma}\label{Lemma4}
Let $(M,g)$ be a space-time and $X$ a global,
nowhere zero nor null, smooth vector field on $M$ (and with associated
covector field denoted by
 $\tilde{X}$) satisfying $\nabla\tilde{X}=\sigma g$ for a smooth function $\sigma$ on $M$.
\begin{enumerate}[(i)]
\item For any $m\in M$ there exists a coordinate neighbourhood $U$ of
$m$ with coordinates functions $u, x^{\alpha}$ ($\alpha,\,\beta=1,2,3$) such
that
 $g$ is given on $U$ by
\begin{equation}\label{36}
ds^{2}=\epsilon du^{2}+\rho ^{2}h_{\alpha \beta}dx^{\alpha}dx^{\beta}
\end{equation}
where $\epsilon =\pm 1$, $h$ is a smooth metric on the level
surfaces of $u$ in $U$ and of signature $(+,+,+)$ for $\epsilon =-1$
and $(-,+,+)$ for $\epsilon =1$, where $\rho$ and $\sigma$ are
smooth functions on $U$ depending only on $u$ and
$\rho^{2}=|g(X,X)|$, $\sigma=d\rho/du$ and
 $X=\rho\partial/\partial u$.

\item If $\sigma$ is a non-zero constant function on $M$ (so that $X$ is homothetic on $M$) then, after
a rescaling of $X$ and a translation of the coordinate $u$, one may
take $\sigma=1$ and (\ref{36}) holds with $\rho=u$.
\end{enumerate}
\end{Lemma}
$\mathbf{Proof}$

The conditions of the theorem show that $X$ is a
nowhere zero nor null conformal vector field (since $\mathcal
L_{X}g=2\sigma g$) with zero conformal bivector and that $\tilde{X}$ is a locally a gradient.
Thus the remarks preceding lemma \ref{Lemma4} show that $X$ is covariantly
constant with respect to the Levi-Civita connection, $\nabla'$, of
$\tilde{g}=\rho^{-2}g$; $\nabla' X=0$. By choosing local coordinates
$t, x^{\alpha}$ in some open neighbourhood $U$ of $m$ one may
arrange that $X^{a}=\partial /\partial t$ and that $\tilde{g}$ is
the local product
\begin{equation}\label{37}
d\tilde{s}^{2}=\epsilon dt^{2}+h_{\alpha \beta}dx^{\alpha}dx^{\beta}
\end{equation}
where $\epsilon (=\pm 1)$ is the sign of $g(X,X)$. The condition
$\nabla \tilde{X}=\sigma g$ shows that, on $U$, $\rho$ and $\sigma$
depend only on $t$. Thus $g=\rho^{2}\tilde{g}$ and a change of the
first coordinate from $t$ to $u$, where $du/dt=\rho$, shows that
$\sigma=d\rho/du$ and reveals (\ref{36}). This establishes part (i)
and part (ii) is immediate from part (i). $\square$

This result leads naturally to the next theorem which is similar to
a result in \cite{6,7}. The proof of this theorem gives a nice
example of a situation where, unlike many of those in the previous section,
non-trivial solutions of the Sinyukov equation (\ref{15}) arise and
how the inversion of the corresponding pair $(a, \lambda)$ to get
the pair $(g', \psi)$ is carried out. This will be useful later this
section.

\begin{Theorem}\label{Theorem6}
Let $(M,g)$ be a space-time and $X$ a global vector field on $M$ with exactly the properties described in the general condition of lemma \ref{Lemma4}. Then the metric $g$ is one of a family of (locally) projectively related metrics on $U$ whose general member is given by $g'=\kappa [Fg-cF^2\tilde{X}\otimes \tilde{X}]$ where $F=(1+\epsilon c\rho^{2})^{-1}$, $\kappa$ and $c$ are constants satisfying the restrictions $\kappa>0$ and $\epsilon c>0$ (and so $F$ is a positive function on $M$) and, as in lemma \ref{Lemma4}, $\epsilon=\pm1$ is the sign of $g(X,X)$. In the coordinates of lemma \ref{Lemma4}(i), $g'$ is given by \begin{equation}\label{38} ds'^{2}=\kappa\{\epsilon F^2du^{2}+F\rho^{2}h_{\alpha \beta}dx^{\alpha} dx^{\beta}\} \end{equation} 
\end{Theorem}
$\mathbf{Proof}$.\\
First note that, under the conditions of the theorem, the tensors $a$ and $\lambda$ defined on $M$ by $a=c_{1}g+c_{2}\tilde{X}\otimes \tilde{X}$ and $\lambda=(c_{2}d\rho/du)\tilde{X}$, where $c_{1}$ and $c_{2}$ are constants satisfying $c_{1}> 0< \epsilon c_{2}$, together satisfy (\ref{15}). Then $a$ is non-degenerate and $\lambda$ is an exact 1-form, on $M$. (It is remarked here that this collection of solutions of (\ref{15}) for the pair $(a, \lambda)$ is not claimed to be the general solution of (\ref{15}) and thus the family of (locally) projectively related metrics on $U$ claimed in the theorem is not claimed to be the complete such family.) 
In appropriate co-ordinates $g$ is given by $\epsilon du^{2}+\rho^2dS^2$ (Lemma \ref{Lemma4}) and in these co-ordinates $a$ is $c_1\left[\epsilon  (1+\epsilon c\rho^2)du^{2}+\rho ^{2}dS^2\right]$ (since $\tilde{X}=\epsilon\rho du$), where $dS^2=h_{\alpha \beta}dx^{\alpha}dx^{\beta}$ and  $c\equiv c_2/c_1$, $\epsilon c>0$. Now (Lemma \ref{Lemma1}) $F=|\det g/\det a|=(1+\epsilon c\rho^2)^{-1}$ and $a^{-1}$, inverse to $a$, on $M$ is easily seen to be $\kappa[\epsilon Fdu^{2}+\rho ^{2}dS^2]$ or $a^{-1}=\kappa[g-cF\tilde{X}\otimes \tilde{X}]$,
where $0<\kappa\equiv1/c_1$. Then $g'=Fa^{-1}$ (Lemma \ref{Lemma1}) gives the result claimed in the theorem.$\square$


\vspace{11pt}

Now consider the general situation when the curvature class of
$(M,g)$ is $\mathbf{C}$ (and so $M$ is non-flat). The only
possibility, for this curvature class, of a local or global metric
$g'$ non-trivially ($\nabla\neq\nabla'$) projectively related to
$g$, wherever both are defined, is when $(M,g)$ has holonomy type
$R_{15}$ (theorem 4(ii)) and when $\lambda$ does not vanish
identically on $M$. So, suppose $m\in M$ with $\lambda(m)\neq 0$ so
that \emph{$\lambda$ is nowhere zero over some connected, open
neighbourhood $V$ of $m$}. (This is where the assumption that the
holonomy type is $R_{15}$ is used.) Suppose, as one can from the
curvature class $\mathbf{C}$ condition, that $V$ is chosen so that
it admits a nowhere-zero smooth (\cite{10}, p262) vector field $k$ such that
$R^{a}{}_{bcd}k^{d}=0$ on $V$ (and with $k(m)$ the unique non-zero
solution of this equation at $m\in V$ up to a scaling). Then the
conditions of lemma \ref{Lemma2} apply and so, on $V$, $\lambda_{a;b}=cg_{ab}$
with $c$ constant (thus $\lambda$ is homothetic) and from (\ref{17}) $\lambda_{a}=\eta k_{a}$ for some smooth
nowhere-zero function $\eta$ on $V$ and, also from (\ref{17}), and theorem 1(ii),
\begin{equation}\label{40}
a_{ab}=\phi g_{ab}+\gamma \lambda_{a}\lambda_{b}
\end{equation}
for smooth functions $\phi$ and $\gamma$ on $V$. Now substitute
(\ref{40}) into (\ref{15}) and, for any $m'\in V$ and for $v\in
T_{m'}M$ which is non-null and ($g$-)orthogonal to $\lambda$ at $m'$
contract with $v^{a}v^{b}$ to see that $\phi_{,a}(m')=0$ and so,
since $V$ is connected, $\phi$ is constant on $V$. A back
substitution and contraction with $v^{a}$ (recalling that $\lambda$
is nowhere zero on $V$) reveals that $\gamma c=1$ on $V$ and hence
that $\gamma =\frac{1}{c}$ is a non-zero constant (and $c\neq 0$).
Thus $(\lambda^{a}\lambda_{a})_{,b}=2c\lambda_{b}$ is nowhere zero
on $V$ from which it follows that $\lambda$, and hence the vector field $k$ which annihilates $Riem$ on $V$ is non-null over some
open dense subset of $V$. Suppose now that $V$ is adjusted (if necessary) so that $\lambda$ is nowhere
zero or null on $V$. Then lemma \ref{Lemma4}(ii) applies for
$\tilde{X}=\lambda$ (after a linear adjustment in the coordinate
$u$) and a family of (non-trivially) projectively related metrics is
given by (\ref{38}).
In this case Lemma \ref{Lemma2}(c) and theorem 1(ii) ensure that (\ref{40}) is the only possible form for $a$, and so the constancy of $\phi$ and $\gamma$ that follows from substitution into (\ref{15}) means that  (\ref{38}) gives {\it all} metrics $g'$ (non-trivially)
locally projectively related (on $V$) to $g$.

\vspace{11pt}

With a choice of timelike co-ordinate $u=t$ (so that $\epsilon=-1$) and $h$ as a 3-metric of constant curvature,
lemma \ref{Lemma4} and theorem 6 apply to the FRWL cosmological metrics. The FRWL cosmological
metrics thus provide examples of non-trivially locally
projectively related pairs of metrics whose common holonomy type is $R_{15}$ and
whose curvature class is, in general, $\mathbf{A}$. However, these
metrics also contain examples where the curvature is of class
$\mathbf{C}$ (and holonomy type $R_{15}$) and are special cases of
the metrics described in theorem 5 of \cite{4}. (The Einstein static
universe is of holonomy type $R_{13}$ and curvature class
$\mathbf{C}$ and generates no non-trivial projectively related
metrics (theorem 4(ii) or 3(ii)).

\vspace{11pt}
Finally, as examples of spacetimes relevant to this section, but where $g'$ of the form (\ref{38}) is {\it not} the general form of the projectively related metrics, consider the spacetime metrics $ds^2=2dudv+u^2d\sigma^2$ (which has holonomy type $R_{11}$) and  $ds^2=\epsilon_1dt^2+\epsilon_2dz^2+z^2d\sigma^2$ (which has holonomy type $R_{10}$ or $R_{13}$, depending on choice of $\epsilon_1=\pm1$ and $\epsilon_2=\pm1$) where $d\sigma^2=h_{\alpha\beta}(x^3,x^4)dx^{\alpha}dx^{\beta}$, each taken as defined on some open subset of $\mathbb{R}^4$ and with the restrictions $u>0$ and $z>0$, respectively. These metrics arise (as equations (7.10) and (7.19)) in \cite{22} as generic examples of curvature class $\mathbf{D}$ spacetimes having non-trivial geodesically equivalent metrics. These spacetimes each admit one global covariantly constant (co)vector field $Y$ and it can be shown that each also admits a global homothetic gradient (co)vector field $X$, so that $Y_{a;b}=0$ and $X_{a;b}=g_{ab}$. The geodesically equivalent metrics $(g',\psi)$ may be derived in each case from the simple construction $a=c_1g+c_2Y\otimes Y+c_3X\otimes X+c_4(X\otimes Y+Y\otimes X), \lambda=c_3X+c_4Y$ where $c_1,c_2,c_3$ and $c_4$ are all real constants.

\subsection{The holonomy types $R_{9}$ and $R_{14}$}

Now suppose that $(M,g)$ is a \emph{non-flat} space-time \emph{of holonomy type $R_{9}$ or $R_{14}$}. Then
$M$ may be disjointly decomposed in an obvious notation as $M=A\cup\mathrm{int}  B\cup\mathrm{int}  C\cup
\mathrm{int}  D\cup F$ (since $\mathrm{int}  O$ is empty) where \cite{10} $A$, $A\cup B$, $A\cup B\cup C$ and
$A\cup B\cup C\cup D$ are open (and the last of these is also dense) in $M$. Thus $F$ is closed and
$M\setminus F$ is open and dense in $M$. Now suppose that $U$ is a non-empty, connected, open subset of $M$ on which is defined a Lorentz metric $g'$ with Levi-Civita connection $\nabla'$ and which is projectively related to $g$ on $U$.
Then one has the disjoint decompositions $U=A'\cup B'\cup C'\cup D'\cup O'=A'\cup\mathrm{int}  B'\cup\mathrm{int}  C'\cup\mathrm{int}  D'\cup F'$ where $A'=A\cap U$, $B'=B\cap U$ etc (and $\mathrm{int}  A'=(\mathrm{int}  A)\cap U=A'$, $\mathrm{int}  B'=(\mathrm{int}  B)\cap U$, etc) $A'$ open in $U$ and $\mathrm{int}  F'=\mathrm{int} (F\cap U)=\emptyset$. Thus
$U\setminus F'$ is open and dense in $U$. If the holonomy type of
$(M,g)$ is $R_{14}$ and any of the subsets $\mathrm{int}  B'$ and $\mathrm{int}  C'$ are
non-empty then, noting that the holonomy algebras associated with
the (necessarily open) components of $\mathrm{int}  B'$ and $\mathrm{int}  C'$ are
subalgebras of the $R_{14}$ holonomy algebra of dimension $\geq 2$ and hence of type
$R_{6}$, $R_{8}$, $R_{9}$, $R_{11}$, $R_{12}$ or $R_{14}$, (for $\mathrm{int} 
C'$) and $R_{7}$ (for $\mathrm{int}  B'$), it follows from theorem 4(ii) and
(iii) applied these components that $\nabla=\nabla'$ on $\mathrm{int}  B'\cup\mathrm{int}  C'$. The situation in $\mathrm{int}  D'$ was described at the end of the last
section. If the holonomy type of $(M,g)$ is $R_{9}$, $B=\emptyset$ and a similar argument using
theorems 4(i) and 4(ii) shows that $\nabla=\nabla'$ on $\mathrm{int}  C'\cup\mathrm{int}  D'$. So interest is directed towards the (assumed non-empty) open subset $A'$ of $U$.

If $W$ is any non-empty, open, connected subset of $A'$ and hence of $U$ it follows from table 1 and the condition $W\subset A'$ that
if $(M,g)$ has holonomy type $R_{9}$ then so also does $(W,g)$ whilst if $(M,g)$ has holonomy type $R_{14}$, $(W,g)$ has holonomy type $R_{9}$, $R_{12}$ or $R_{14}$. If the metrics $g$ and $g'$ are projectively related on $U$ and if the associated 1-form $\lambda$ vanishes on $W$, (\ref{15}) shows that $\nabla a=0$ on $W$. Since $W\subset A'$ it follows from theorem 1(iv) and the remarks following it that $a=\alpha g$ on $W$ ($\alpha\in \mathbb{R}$). Theorem 3(iii) then shows that the trivial solution $\lambda=0$ and $a=\alpha g$ on $U$ is the only solution of (\ref{15}) on $U$ and hence
$\lambda=0$ and $a=\alpha g$ is the only solution of (\ref{15}) on $M$. So suppose that $\lambda$ does not vanish on any non-empty, open subset of $A'$. Then the subset $A''\equiv A'\setminus \{m\in A':\lambda(m)=0\}$ is open and dense in $A'$ and open in $U$. Now, with $(M,g)$ of holonomy type $R_{9}$ or $R_{14}$, for any non-empty open subset $W\subset A''$, $(W,g)$ has holonomy type $R_{9}$ or $R_{14}$ (since, if it were $R_{12}$, $\lambda\equiv 0$ on $W$ from \cite{22} or theorem 3(i)) and so each $m\in A''$ admits a connected, open neighbourhood $V\subset A''$ on which is defined a nowhere-zero ($\nabla$-)recurrent null vector field $l$ which is unique up to a nowhere-zero scaling. The recurrence condition on $l$
shows that $l$ is hypersurface-orthogonal on $V$ (so with all index movements done using the metric $g$, $l_{[a;b}l_{c]}=0$) and so $V$ may be chosen so that a (nowhere-zero) rescaled version of $l$ (also denoted $l$) exists on it and which is (recurrent and) normal, satisfying $l_{a}=u_{,a}$ and, from the recurrence condition, $l_{a;b}=\beta l_{a}l_{b}$ for functions $u$ and $\beta$ on $V$. The set $V$ may be chosen so that the restriction of $l$ to $V$ may be augmented into a smooth null tetrad $l,n,x,y$ on $V$. Then the Ricci identity for $l$ gives on $V$ (again using a comma for a partial derivative)
\begin{equation}\label{41}
(l_{a;b}=\beta l_{a}l_{b})\Rightarrow
l_{d}R^{d}{}_{abc}=l_{a;bc}-l_{a;cb}=l_{a}F_{bc}\qquad
(F_{ab}=2l_{[a}\beta_{,b]})
\end{equation}
Since there are no non-trivial solutions for $k$ of the equation
$k^{a}R_{abcd}=0$ at any point of
$A$ and hence of $A''$, $F$ and hence $d\beta$ are nowhere zero on $V$.
Now, from
(\ref{41}), $l\wedge x$ and $l\wedge y$ lie in the kernel of the
map $f$ in (\ref{1}) at each point of $V$ and so from lemma \ref{Lemma2}, $l$, $x$ and $y$ are
eigenvectors of $\nabla \lambda$ with equal eigenvalues. Thus
\begin{equation}\label{42}
\lambda_{ab}\equiv \lambda_{a;b}=\rho g_{ab}+\sigma l_{a}l_{b}
\end{equation}
holds on $V$ for functions $\rho$ and $\sigma$ which are smooth on
$V$ since $\lambda_{a;b}n^{a}n^{b}$ and $\lambda^{a}{}_{a}$ are.
Then using (\ref{42}) in (\ref{16}) one gets
\begin{equation}\label{43}
a_{ae}R^{e}{}_{bcd}+a_{be}R^{e}{}_{acd}=\sigma
(g_{ac}l_{b}l_{d}+g_{bc}l_{a}l_{d}-g_{ad}l_{b}l_{c}-g_{bd}l_{a}l_{c})
\end{equation}
A contraction of (\ref{43}) with $l^{a}$ (defining
$L_{a}=a_{ab}l^{b}$ and noting that it is nowhere-zero on $V$ since
$a$ is non-degenerate on $V$) and use of (\ref{41}) gives
$L_{e}R^{e}{}_{bcd}=L_{b}F_{cd}$ on $V$. Then a contraction of this
equation first with $L^{b}$ shows that $L$ is null and second with
$l^{b}$ (and a comparison with (\ref{41}) contracted with $L^{a}$)
shows that $L^{e}l_{e}F_{ab}=0$ and hence that $L_{a}l^{a}=0$, on $V$. It
follows that $L_{a}$ and $l_{a}$ are proportional on $V$ and so
\begin{equation}\label{44}
a_{ab}l^b=\phi l_a
\end{equation}
for some clearly smooth function $\phi$ on $V$ which is nowhere zero
on $V$ because of the non-degeneracy of $a$.

Since $l=du$ on $V$, a differentiation
of (\ref{44}) and use of (\ref{15}) (and $l_{a;b}=\beta
l_{a}l_{b}$) gives
\begin{equation}\label{45}
g_{ab}(\lambda_cl^c)+\lambda_al_b=l_a\phi_{,b} \qquad(\Rightarrow \ \
\ \lambda_a=\phi_{,a}=\phi'l_{a})
\end{equation}
where the last steps follow from a rank argument and the fact that $l$ is nowhere zero on $V$ (to get
$\lambda_{a}l_{b}=l_{a}\phi_{,b}$) and the fact that $\phi$ is
a (nowhere-zero) function of $u$ only (and a prime denotes $d/du$). Thus
$\phi'$ is a smooth nowhere-zero function of $u$ on $V$. So the function $\rho$ in (\ref{42}) vanishes on $V$. It is then noted,
from section 4, that $\phi$, being a potential of $\lambda$,
satisfies $\phi+c'=\frac{1}{2}a_{ab}g^{ab}$ for some $c'\in
\mathbb{R}$. It follows that $\lambda$ may be taken as representative of the recurrent null direction on $A''$ and one may take $l=\lambda$ on each such neighbourhood $V$. Then the function $\beta$ may be redefined as that associated with $\lambda$ ($\lambda_{a;b}=\beta \lambda_{a} \lambda_{b}$) on each such $V$. Now $\beta$ is defined on $A''$ and it easily follows from (\ref{41}) that if $\beta$ vanishes over any non-empty open subset of $A''$ the Ricci identity contradicts the curvature class $\mathbf{A}$ condition on that set. Hence, for later convenience, one may remove the set of zeros of $\beta$ from $A''$ leaving behind an open dense subset $A'''$ of $A''$ (and hence of $A'$) on which $\beta$ is nowhere-zero. It also follows from (\ref{45}) that, with this identification of $l$ and $\lambda$, that $\phi$ may, up to the choice of additive constant, be chosen as the co-ordinate $u$. This will be done at a convenient point later in the argument

Continuing the argument one has $\lambda_{a;b}=\beta \lambda_{a}\lambda_{b}$ on $V$ and so, from (\ref{44}),
$a_{ab}l^b=\phi l_b$.
Now contract (\ref{43}) with $l^{c}$ and use (\ref{41}) to get
\begin{equation}\label{48}
\phi(l_a\beta_{b}+\beta_{a}l_b)-(l_aa_{be}\beta^e+a_{ae}\beta^el_b)=2\beta
l_al_b
\end{equation}
where $\beta^{a}=g^{ae}\beta_{e}$ and $\beta_{a}\equiv \beta_{,a}$.
A contraction of (\ref{48}) with $n^{b}$ and then another with $n^{a}n^{b}$ and some algebraic
simplifications then yields
\begin{equation}\label{49}
a_{ab}\beta^b=\phi\beta_a-\beta l_a
\end{equation}

Now the choice $l=\lambda$ means that $l$ and $\beta$ are defined, smooth and nowhere-zero on $A'''$. So consider the function $\beta_{,a}l^{a}$ and note that it cannot vanish over any non-empty open subset $\bar{V}$ of $A'''$.
For if it did then
$(\beta_{,a}l^{a})_{;b}=0$ on $\bar{V}$ and so, by the recurrence of $l$,
$l^{a}\beta_{a;b}=0$ on $\bar{V}$. But then a differentiation of
(\ref{49}) followed by a contraction with $\beta^{a}$ and use of
(\ref{15}) and (\ref{49}) (noting that, on $\bar{V}$,
$\beta^{a}\lambda_{a}=\beta^{a}l_{a}=0$ from (\ref{45}) and that, from (\ref{49})
$a_{ab}\beta^{a}\beta^{b}_{;c}=\phi\beta_{b}\beta^{b}_{;c}$) shows
that $\beta_{a}$ is null in addition to satisfying
$\beta_{a}l^{a}=0$ on $\bar{V}$. Thus $\beta_{a}$ is proportional to
$l_{a}$ and the bivector $F$ in (\ref{41}) vanishes on $\bar{V}$. This
contradiction completes the argument. So a further reduction to an open dense subset $\tilde{A}$ of $A'''$(and hence of $A'$) will be made by omitting those points where $\beta_{,a}l^{a}$ vanishes. Henceforth, one works on $\tilde{A}$.

It follows that the bivector $F$ is
timelike over $\tilde{A}$ and one may choose a connected, open neighbourhood $V\subset \tilde{A}$ about any $m\in \tilde{A}$  and a smooth null tetrad $l,n,x,y$ on $V$ so that, on $V$,
\begin{equation}\label{50}
\beta_a=p n_a+ql_a
\end{equation}
for smooth functions $p$ and $q$ on $V$ with $p$ ($=\beta_{a}l^{a}$)
nowhere zero on $V$ and $F$ is now the bivector $2pl_{[a}n_{b]}$. On
substituting (\ref{50}) into (\ref{49}) one finds

\begin{equation}\label{51}
a_{ab}n^b=\phi n_a+\xi l_a, \ \ \ \ \ \xi\equiv-\beta/p
\end{equation}
so that $\xi$ is smooth and nowhere-zero on $V$ (since $\beta$ is). Equations (\ref{44}) and (\ref{51}) reveal that the 2-space,
$l\wedge n$, is, at each $m\in V$, a timelike \emph{invariant} 2-space of $a$
and so  it follows that the orthogonal complement, $x\wedge
y$, of $l\wedge n$ is also (see \cite{10}). Since this latter 2-space is spacelike
it follows that, at each $m\in V$,
the spacelike vectors $x$ and $y$ may be adjusted so that they are
eigenvectors of $a$ at $m$ \cite{10}. However, such an adjustment may not lead
to \emph{smooth} eigenvector fields of $a$ over some open
neighbourhood of $m$ \cite{24}. To consider this problem further let
a smooth second order symmetric tensor field $K$ be defined on $V$
in the original smooth (unadjusted) tetrad $l,n,x,y$ by

\begin{equation}\label{52}
K_{ab}=a_{ab}-\phi(l_{a}n_{b}+n_{a}l_{b})-\xi l_{a}l_{b}
\end{equation}
Then $K_{ab}l^{b}=K_{ab}n^{b}=0$ on $V$ and $x\wedge y$ is clearly an
invariant 2-space of $K$ at each $m\in V$ and which contains an
orthogonal pair of eigenvectors $\tilde{x}$ and $\tilde{y}$ of $K$
but which may not lead to smooth eigenvector fields of $K$ on $V$
(but the 2-spaces, $x\wedge y$ and $\tilde{x}\wedge \tilde{y}$
coincide). Using the tetrad $l,n,\tilde{x},\tilde{y}$ at $m$ one has
$K_{ab}\tilde{x}^{b}=\tilde{C}g_{ab}\tilde{x}^{b}\equiv \tilde{C}\tilde{x}_{a}$
and similarly, $K_{ab}\tilde{y}^{b}=\tilde{D}\tilde{y}_{a}$ for
$\tilde{C}, \tilde{D} \in R$. Then, from (\ref{52})
$a_{ab}\tilde{x}^{b}=\tilde{C}\tilde{x}_{a}$ and
$a_{ab}\tilde{y}^{b}=\tilde{D}\tilde{y}_{a}$ at $m$ and
$\tilde{C}\neq 0\neq \tilde{D}$ since $a$ is non-degenerate. It
follows that the Segre type of $K$ at a given point of $V$ is either
$\{(1,1)11\}$ (if $\tilde{C}\neq \tilde{D}$ at that point) or
$\{(1,1)(11)\}$ (if ($\tilde{C}=\tilde{D}$ at that point) and no
further degeneracies are permitted (since the repeated eigenvalue in
each 2-space, $l\wedge n$, is zero) \cite{10}. Since this is true for each such set $V$, it is true over $\tilde{A}$ and so
$\tilde{A}$ may be written as the disjoint union of two (not necessarily non-empty) sets, upon each (non-empty) one of which the Segre type of $K$ (including degeneracies) is constant. The member $A_{1}$ of this pair of sets corresponding to Segre type $\{(1,1)11\}$ is then open in $\tilde{A}$ (and hence in $A'$) since the solutions (eigenvalues) of the characteristic polynomial of $a$ depend smoothly on the (smooth) coefficients of this polynomial. If the set corresponding to Segre type $\{(1,1)(11)\}$ is denoted by $A_{2}$ then $A_{1}\cup\mathrm{int}  A_{2}$ is an \emph{open dense} subset of $\tilde{A}$ and hence of the original $A'$ and all eigenvalues of $K$ and hence of $a$ are smooth on $A_{1}$ and $\mathrm{int}  A_{2}$ since the Segre type is constant on each of them. The associated eigenvector fields may then be chosen locally smoothly (in some simply connected open neighbourhood of any point of $A_{1}$ and $\mathrm{int}  A_{2}$) \cite{24}.
Thus one has an open dense subset of $A'$ any point of which admits a neighbourhood with all properties so far derived for such sets and additionally on which a
null tetrad $(l,n,x,y)$ may be chosen with $l$ and $n$
as before, so that (\ref{52}) still holds and
$K_{ab}=Cx_{a}x_{b}+Dy_{a}y_{b}$ with \emph{smooth} eigenvector
fields now labelled $x$ and $y$ and \emph{smooth} eigenvalues now labelled $C$
and $D$ and with $C$ and $D$ everywhere distinct (respectively,
everywhere equal) if the Segre type of $K$ is $\{(1,1)11\}$
(respectively $\{(1,1)(11)\}$). Thus, on any such neighbourhood,
(\ref{52}) gives $a_{ab}x^{b}=Cx_{a}$ and $a_{ab}y^{b}=Dy_{a}$.

Henceforth, attention will be directed to the two open sets $A_{1}$ and $\mathrm{int}  A_{2}$.
These open sets can be related to the holonomy structure and can be described (briefly) in the following way. Consider the range space of the map $f$ in (\ref{1}) at $m\in \tilde{A}$.
It is clear from (\ref{41}) that if a bivector
$G$ is in the range of $f$ at $m$ (so that for some bivector $H$,
$G_{ab}=R_{abcd}H^{cd}$ at $m$) then $G_{ab}l^{b}=\mu l_{a}$ for
$\mu \in \mathbb{R}$. In terms of the null tetrad fields constructed just before (\ref{50}) this means
that the range space of $f$ at $m$ is spanned by the bivectors
$l\wedge n$, $l\wedge x$, $l\wedge y$ and $x\wedge y$ (subject to
consistency with the curvature class $\mathbf{A}$ condition). Now,
recalling the Ambrose-Singer theorem (see the end of section 3), the
holonomy algebra of $(W,g)$ for some connected open neighbourhood $W\subset \tilde{A}$ of $m$ can be constructed by first choosing (any) $m\in W$ and, for each $m'\in U$ and each curve $c$ from $m'$
to $m$ in $W$, computing the range space of $f$ at $m'$ and parallely
transporting each member of it to $m$ along $c$. If this is done for
each such $m'$ and $c$, the collection of bivectors accumulated at
$m$ spans the holonomy algebra of $(W,g)$. Suppose that the algebra
thus arising contains a member which does \emph{not} satisfy the condition that it is a simple bivector whose
blade contains $l$ (and this must be the case if the holonomy type
of $(W,g)$ is $R_{14}$). Now because $l$ is recurrent, it gives
rise, by parallel transport, to a 1-dimensional (null) holonomy
invariant distribution on $W$. Thus any bivector in the
range space of $f$ at some point of $W$ will, if it is simple with
$l$ in its blade, remain that way under parallel transport. It
follows that the range space of $f$ at some $m'\in U$ must contain a
bivector which does not have the property that it is simple with $l$ in its blade. From this it
follows that $R_{abcd}x^{[c}y^{d]}\neq 0$ at $m'$ and hence in
some connected open neighbourhood of $m'$.
If all bivectors in the holonomy algebra of $(W,g)$ (and hence in
the range space of $f$ at any $m'\in U$) are simple with $l$ in
their blade the holonomy group of $(W,g)$ is $R_{9}$ and
$R_{abcd}x^{[c}y^{d]}=0$ at each point of $W$.

Now contract (\ref{43}) with $x^{a}y^{b}$ and use the results
$a_{ab}x^{b}=Cx_{a}$ and
$a_{ab}y^{b}=Dy_{a}$
to get, at any point of $\tilde{A}$
\begin{equation}\label{53}
(C-D)R_{abcd}x^{[c}y^{d]}=0
\end{equation}
Thus if the holonomy group of $(W,g)$ is of type $R_{14}$ the above work together with (\ref{53}) shows that $W$ must intersect $A_{2}$ (and, in fact, $\mathrm{int}  A_{2}$) and hence that $m$ lies in the \emph{closure} of $A_{2}$.
Let $m\in A_{1}\cup\mathrm{int}  A_{2}$ and let $V$ be a connected, open neighbourhood of $m$ with the properties given to such neighbourhoods just before (\ref{50}) and which is contained in exactly one of these two (disjoint) sets. Then $C$ and $D$ may be taken as smooth eigenvalues of $K$ on $V$ with associated smooth eigenvectors $x$ and $y$. If the holonomy type of $(V,g)$ is $R_{14}$,
$V\subset A_{2}$ and $C=D$ on $V$, whilst if the holonomy type of $(V,g)$ is $R_{9}$, one could have either $V\subset A_{2}$ and $C=D$ on $V$ or $V\subset A_{1}$ and $C\neq D$ on $V$.
Considering the holonomy type $R_{14}$ first, (\ref{52}) gives
\begin{eqnarray}\label{54}
a_{ab}=\phi(l_an_b+n_al_b)+\xi
l_al_b+C(x_ax_b+y_ay_b)\\\nonumber=Cg_{ab}+\xi
l_{a}l_{b}+(\phi-C)(l_an_b+n_al_b)
\end{eqnarray}
where the completeness relation
$g_{ab}=l_{a}n_{b}+n_{a}l_{b}+x_{a}x_{b}+y_{a}y_{b}$ has been
employed. Since $\tfrac{1}{2}a_{ab}g^{ab}=(\phi+C)$ and $\phi$
and $u$ are each potentials for $\lambda=l$, \emph{$C$ is constant on $V$}.
Then if we choose to adapt coordinates to suit this particular
$(a,\lambda)$ solution one can choose the zero of coordinate $u$
such that $\phi=u+C$, and (\ref{54}) becomes
\begin{equation}\label{55}
a_{ab}=u(l_an_b+n_al_b)+\xi l_al_b+Cg_{ab}
\end{equation}

If the holonomy group is of type $R_{9}$ then there is no control
over $C$ and $D$ and one may write from (\ref{52}) (to cover either
of the possibilities)

\begin{equation}\label{56}
a_{ab}=u(l_an_b+n_al_b)+\xi l_al_b+C(x_ax_b +y_ay_b)+Ey_{a}y_{b}
\end{equation}
 where $E\equiv D-C$ is smooth on $V$.

With $V$ as above, suppose that the holonomy type of $(V,g)$ is
$R_{14}$ so that (\ref{55}) holds with $C$ constant. Then a
substitution of (\ref{55}) into (\ref{15})
(using $l_{a.b}=\beta l_{a}l_{b}$ and $n^{a}n_{a;b}=0$) and a
contraction first with $n^{a}n^{b}$ gives

\begin{equation}\label{57}
2n_a-2\beta\xi l_a=\xi_{,a}
\end{equation}
and then with $n^{a}$ gives

\begin{equation}\label{58}
un_{a;b}=(g_{ab}-l_an_b-n_al_b)-u\beta n_al_b
\end{equation}
Then (\ref{58}) gives an expression for $n_{a;b}$ from which it is clear from the non-degeneracy of $g$ that $u$ cannot vanish on $V$ and
easily checked that $n_{[a;b}n_{c]}=0$ on $V$ and so $n$ is
\emph{hypersurface-orthogonal} on $V$. Thus reducing $V$, if necessary, some nowhere-zero scaling $\tilde{n}$ of $n$ satisfies
$\tilde{n}=dv$ for some smooth function $v$ on $V$.  Then from (\ref{58}) and the recurrence of $l$ it is easily seen that $l$ and $n$ are involutive
and so span a 2-dimensional smooth distribution $D_{1}$ on $V$. Also the subsets of $V$ of constant $u$ \emph{and} $v$ span a smooth 2-dimensional distribution $D_{2}$ on $V$. Now, for $m\in V$, $T_{m}M = D_{1}(m)\oplus D_{2}(m)$ and so by considering the local flat charts for $D_{1}$ and $D_{2}$, respectively, one may choose $V$ and a coordinate system $y^{a}$ on $V$ such that $\{\partial/\partial y^{1},\partial/\partial y^{2}\}$ span $D_{1}$ and $\{\partial/\partial y^{3},\partial/\partial y^{4}\}$ span $D_{2}$, on $V$ (see e.g. \cite{Spivak} chapter 6, page 28). One can then check by reintroducing the functions $u$ and $v$ and relabelling $y$ by $x$ that $V$ may be chosen as a coordinate domain with coordinate functions $u$, $v$, $x^{3}$, $x^{4}$ such that the metric $g$ takes the form
\begin{equation}\label{59}
ds^2=2Pdudv+g_{\alpha\beta}dx^{\alpha}dx^{\beta}
\end{equation}
where $P$ is a smooth function on $V$ and Greek letters take the
values 3, and 4. [Thus in this coordinate system, $l_{a}$ has
components $(1,0,0,0)$, $l^{a}=(0,P^{-1},0,0)$, $n_{a}=(0,P,0,0)$
and $n^{a}=(1,0,0,0)$.] On writing out the recurrence condition on
$l$ in these coordinates as $l_{c}\Gamma ^{c}_{ab}=-\beta
l_{a}l_{b}$ one easily finds that $P$ is independent of $x$ and $y$
and that the $g_{\alpha \beta}$ are independent of $v$. Then
(\ref{58}) in these coordinates (and with $a=\alpha, b=\beta$) gives
$-un_{c}\Gamma ^{c}_{\alpha \beta}=g_{\alpha \beta}$ which is
$-uP\Gamma^{2}_{\alpha \beta}=g_{\alpha \beta}$ and which easily
leads to $u\partial g_{\alpha \beta}/\partial u=2g_{\alpha \beta}$.
Thus, after possibly a change of coordinates $x^{3}$ and $x^{4}$
(and possibly also of the open subset $V$) the metric becomes

\begin{equation}\label{60}
ds^2=2P(u,v)dudv+
u^2e^{2w(x^{3},x^{4})}((dx^{3})^2+(dx^{4})^2)
\end{equation}
for some smooth function $w$ on $V$. Further remarks must be added
here regarding the function $w$ and this is, perhaps, best done
after the holonomy type $R_{9}$ has been considered.

Suppose now that the holonomy type of $(V,g)$ is $R_{9}$. Then
either (\ref{55}) or (\ref{56}) may hold on $V$ with the
nowhere-zero functions $C$ and $D$ and the vector fields $x$ and $y$
smooth on $V$ and if it is (\ref{55}) one again arrives at
(\ref{60}). If (\ref{56}) holds with $C$ and $E$ smooth, distinct
and nowhere-zero on $V$ then $C=a_{ab}x^{a}x^{b}$ and
$C+E=a_{ab}y^{a}y^{b}$ and differentiating, using (\ref{15}), $l=\lambda$ and $x_{a}x^{a}_{;b}=y_{a}y^{a}_{;b}=0$, one easily finds that
\emph{$C$ and $E$ are distinct non-zero constants on $V$}. Now write
$a_{ab}x^{b}=Cx_{a}$, differentiate and contract with $y^{a}$ using
using (\ref{15}) and (\ref{56}). One finds that $Ey^{a}x_{a;b}=0$.
Similarly one can get $Ex^{a}y_{a;b}=0$ It follows that
$y^{a}x_{a;b}=0$ and $x^{a}y_{a;b}=0$ on $V$. Since
$l^{a}x_{a;b}=l^{a}y_{a;b}=0$ from the constancy of $l^{a}x_{a}$ and
$l^{a}y_{a}$, respectively, and the recurrence condition on $l$, one
gets $x_{a;b}=l_{a}r_{b}$ and $y_{a;b}=l_{a}s_{b}$ for smooth 1-form
fields $r$ and $s$ on $V$. It is then easily checked, from a back
substitution, that $x_{a}=-ur_{a}$ and $y_{a}=(E-u)s_{a}$ and so, since then ($E-u$) cannot vanish on $V$,
\begin{equation}\label{61}
x_{a;b}=-u^{-1}l_{a}x_{b}\qquad
y_{a;b}=(E-u)^{-1}l_{a}y_{b}
\end{equation}
holds on $V$. Equation (\ref{61}) shows that the vector fields $x$
and $y$ are hypersurface orthogonal;  $x_{[a;b}x_{c]}=0$,
$y_{[a;b}y_{c]}=0$. The calculation for $n_{a;b}$ which produced
(\ref{58}) in the case $C=D$ goes through in a similar way to
before, and is now

\begin{equation}\label{62}
un_{a;b}=(g_{ab}-l_an_b-n_al_b)-u\beta n_al_b+(\frac{E}{u-E})y_{a}y_{b}
\end{equation}
However, the consequence $n_{[a;b}n_{c]}=0$ still holds and, again,
$n$ is hypersurface orthogonal. One thus again arrives at the metric
(\ref{59}). But now the hypersurface orthogonality of $x$ and $y$
means that $V$ may be chosen so that one may write $x_{a}=hq_{,a}$
and $y_{a}=h'q'_{,a}$ for smooth functions $h$, $h'$, $q$ and $q'$
on $V$. Then using the coordinates $u, v, x^{3}(=q)$ and
$x^{4}(=q')$, write out (\ref{62}) with the indices $a$ and $b$ set equal to $\alpha$ and $\beta$ as
before, to get $-un_{c}\Gamma ^{c}_{\alpha \beta}=g_{\alpha
\beta}+E(u-E)^{-1}y_{\alpha}y_{\beta}$. The only new information is
yielded by the choices $\alpha=\beta=3$ and $\alpha=\beta=4$ and is
$g_{33}=G_{33}(x^{3},x^{4})u^{2}$ and
$g_{44}=G_{44}(x^{3},x^{4})(u-E)^{2}$ for smooth functions $G_{33}$
and $G_{44}$ on $V$. A back substitution into either of (\ref{61})
then shows that $G_{33}$ is independent of $x^{4}$ and that $G_{44}$ is
independent of $x^{3}$. Thus the functions $G_{33}$ and $G_{44}$ may be
absorbed into the coordinates and in these new coordinates, $x$ and $y$, on open
subset $V$ the metric becomes
\begin{equation}\label{63}
ds^2=2P(u,v)dudv+ u^2dx^2+(u-E)^2dy^2
\end{equation}
For either metric (\ref{60}) or (\ref{63}) the final piece of
information, that is, the nature of the function $P(u,v)$, may be deduced from (\ref{57}). First note from (\ref{51}) and (\ref{57}) that the function $\xi$ is a nowhere-zero
function of $u$ and $v$ only, on $V$. Then the recurrence condition on
$l$ and either (\ref{60}) or (\ref{63}) give $\beta=-P^{-1}\partial
P/\partial u$ from which (\ref{57}) gives

\begin{equation}\label{64}
\tfrac{\partial\xi}{\partial u}=2\xi\tfrac{1}{P}\tfrac{\partial
P}{\partial u}\qquad \tfrac{\partial\xi}{\partial v}=2P\qquad
\tfrac{\partial\xi}{\partial x}=0\qquad\tfrac{\partial\xi}{\partial
y}=0
\end{equation}

The first equation in (\ref{64}) gives $P=\sqrt{\xi}\tfrac{df}{dv}$ where $f$ is some arbitrary function of only $v$. The second equation in (\ref{64}) then becomes
$\tfrac{\partial\xi}{\partial v}=2\sqrt{\xi}\tfrac{df}{dv}$, whose solution is $\xi=(f(v)+B(u))^2$ where $B$ is some arbitrary function of only $u$. Thus we have
$\tfrac{\partial\xi}{\partial u}=2\sqrt{\xi}\frac{dB}{du}$.
Now note that, from (\ref{64}), $2Pdudv=\tfrac{\partial\xi}{\partial v}dudv$. A change of coordinates from $(u,v)$ to $(u,z=\tfrac{1}{2}\xi)$ gives
$2Pdudv=2dudz-4\tfrac{\partial\xi}{\partial u}du^2$ and hence that, with $b(u)$ an arbitrary function, $2Pdudv$ may be expressed in the form $2dudz+\sqrt{z}b(u)du^2$.

In summary, if $(V,g)$ has holonomy type $R_{14}$, $C (=D$) is a
non-zero constant and the metric takes the form
\begin{equation}\label{66}
ds^2=2dudz+\sqrt{z}b(u)du^2+u^2e^{2w(x^{3},x^{4})}((dx^{3})^2+(dx^{4})^2)
\end{equation}
for some function $w$ whereas
if $(V,g)$ has holonomy type $R_{9}$ the metric takes either the
form (\ref{66}) or
\begin{equation}\label{67}
ds^2=2dudz+\sqrt{z}b(u)du^2+ u^2dx^2+(u-E)^2dy^2.
\end{equation}
To distinguish between
the holonomy types for the metric (\ref{66}), one may calculate that
the extra bivectors required in the range of the map $f$ in
(\ref{1}) (and alluded to just before (\ref{53})) to exclude the
$R_{9}$ case will be zero on $V$ if the function $w$ satisfies the
harmonic condition $\partial ^{2}w/\partial (x^{3})^{2}+\partial
^{2}w/\partial (x^{4})^{2}=0$ on $V$. Thus the metric (\ref{66}) is of
holonomy type $R_{9}$ if and only if $w$ is harmonic over $V$.

The general form for the solution pairs $(a,\lambda)$ of (\ref{15}) and hence of solution pairs $(g',\psi)$ of (\ref{11}) can now be found.
For (\ref{66}) one may calculate that $n_a=(\tfrac{1}{2}\sqrt{z}b(u),1,0,0)$ and that $\xi=2z$. 
It can be shown that \cite{Hall&LonieSigma} because of a uniqueness in the direction of $\lambda$ the only freedom in $a$ is that of a constant scaling and the addition of a constant multiple of the metric. Hence the most general form of $a$ may be written in the form
\begin{equation}\label{68}
a_{ab}=\kappa(cu(l_{a}n_{b}+n_{a}l_{b})+2czl_{a}l_{b}+g_{ab}), \ \ \
\ \lambda_{a}=c\kappa l_{a}
\end{equation}
where $c$ and $\kappa>0$ are constants, and with $u$ restricted in the chart domain to ensure that $a$ is non-degenerate, i.e. that $1+cu\neq0$.
Inverting one finds
\begin{equation}\label{69}
F=e^{2\chi}=\frac{1}{\kappa^4(1+cu)^2}
\end{equation}
and then after calculating $a^{-1}$, $g'=Fa^{-1}$ is of the form
\begin{equation}\label{70}
\frac{1}{\kappa^5}\left(\frac{2dudz}{(1+cu)^3}+\frac{(\sqrt{z}b(u)(1+cu)+2cz)du^2}{(1+cu)^4}
+\frac{u^2e^{2w(x^{3},x^{4})}((dx^{3})^2+(dx^{4})^2)}{(1+cu)^2}\right)
\end{equation}
The associated 1-form is $\psi=d(\tfrac{1}{2}\ln F)$, given by
\begin{equation}\label{71}
\psi_{a}=-\frac{c}{1+cu}l_{a}
\end{equation}
It is possible to express (\ref{70}) in a simplified form (identical in form to $g$) through a co-ordinate transformation from $u$ and $z$ to $U$ and $Z$ where $(u=U/(1-cU),z=Z/(1-cU))$ to obtain, with $B(U)\equiv b(U/(1-cU))/(1-cU)$
\begin{equation}\label{75}
\frac{1}{\kappa^5}\left(2dUdZ+\sqrt{Z}B(U)dU^2+U^2e^{2w(x^{3},x^{4})}((dx^{3})^2+(dx^{4})^2)\right)
\end{equation}

And similarly, for (\ref{67})
\begin{equation}\label{72}
a_{ab}=\kappa(cu(l_{a}n_{b}+n_{a}l_{b})+2czl_{a}l_{b}+g_{ab}+cEy_{a}y_{b}),
\ \ \ \ \lambda_{a}=c\kappa l_{a}
\end{equation}
where $c$, $E$ and $\kappa>0$ are constants, and with $u$ restricted in the chart domain to ensure $1+cu\neq0$, and with $c$ and $E$ chosen such that $1+cE\neq0$, so that $a$ is non-degenerate.
Inverting one finds, 
\begin{equation}\label{73}
F=e^{2\chi}=\frac{1}{\kappa^4(1+cu)^2(1+cE)}
\end{equation}
and again, after calculating $a^{-1}$,  $g'=Fa^{-1}$ takes the form
\begin{equation}\label{74}
\frac{1}{\kappa^5(1+cE)}\left(\frac{2dudz}{(1+cu)^3}+\frac{(\sqrt{z}b(u)(1+cu)-2cz)du^2}{(1+cu)^4}+\frac{u^2dx^2}{(1+cu)^2}+\frac{(u-E)^2dy^2}{(1+cE)(1+cu)^2}\right)
\end{equation}
The projective 1-form $\psi$ has the from given in (\ref{71}).
As in the previous case a co-ordinate transformation, in this case $(u=\mu U/(1-c\mu U),z=\mu Z/(1-c\mu U),y=Y/\mu)$, allows (\ref{74}) to be expressed in a simplified form 
\begin{equation}\label{75}
\frac{1}{\kappa^5}\left(2dUdZ+\sqrt{Z}B(U)dU^2+U^2dx^2+\left( U-E'\right)^2dY^2)\right)
\end{equation}
where $\mu=\sqrt{1+cE}$, $B(U)\equiv1/(1-c\mu U)b\left(\mu U/(1-c\mu U)\right)$  and $E'=E\mu^{-3}$.


Thus if $(M,g)$ is non-flat  and locally projectively related over a non-empty connected open subset $U$ to $(U,g')$ then, in the previous notation,  if $(M,g)$ is of holonomy type $R_{9}$  each point of an open dense subset of $U$ admits an open neighbourhood $V$ on which either $\nabla'=\nabla$ or $g$ and $g'$ satisfy (\ref{66}) and (\ref{70}) (with $w$ harmonic on $V$) or (\ref{67}) and (\ref{74}) whereas  if $(M,g)$ is of holonomy type $R_{14}$  each point of an open dense subset of $U$ admits an open neighbourhood $V$ on which either $\nabla'=\nabla$ or $g$ and $g'$ satisfy the conditions mentioned earlier for the holonomy type $R_{11}$ case (as dealt with in \cite{22}) or $g$ and $g'$ satisfy  (\ref{66}) and (\ref{70}) .

\section{Acknowledgements}
The authors thank Wang Zhixiang for several useful discussions.

\end{document}